\documentclass[twocolumn,3p,times,procedia]{elsarticle}
\usepackage{ecrc}
\usepackage{graphicx}
\usepackage{titlesec}
\usepackage{array,threeparttable}
\usepackage{url}
\usepackage{amsthm}
\usepackage{amsmath,amssymb,amsfonts}
\usepackage{bm}
\usepackage{algorithm}
\usepackage{algorithmic}
\usepackage{graphicx}
\usepackage{subfigure}
\usepackage{epsfig,latexsym}
\usepackage{textcomp}
\usepackage{xcolor}
\usepackage{epstopdf}
\usepackage{stfloats}
\usepackage{makecell}
\usepackage{tabularx}
\usepackage{multirow}
\usepackage{booktabs}

\volume{00}

\newtheorem{proposition}{Proposition}

\newtheorem{corollary}{Corollary}

\firstpage{1}
\runauth{}
\jid{procs}
\usepackage{amssymb}
\usepackage{amsmath}
\usepackage{multicol}
\usepackage[figuresright]{rotating}
\usepackage{bm}
\usepackage{caption}
\usepackage{fancyhdr}

\fancyheadoffset[RO,EL]{0pt}
\usepackage{geometry}

\begin{document}

\begin{frontmatter}

\dochead{}
\title{
\begin{flushleft}
{\bf \Huge Towards 6G Native-AI Edge Networks: A Semantic-Aware and Agentic Intelligence Paradigm} 
\end{flushleft}
}

\author[]{\bf \Large \leftline {Chenyuan Feng$^1$,  Anbang Zhang$^2$, Geyong Min$^1$, Yongming Huang$^{3,4}$,}
\bf \Large \leftline {Tony Q. S. Quek$^{4,5,*}$, Xiaohu You$^{3,4,*}$}}

\address{
\bf  \leftline {$^1$ College of Computer Science, University of Exeter, Exeter, U.K. }

\bf  \leftline {$^2$ Sony China Research Laboratory, Beijing, China} 


\bf  \leftline {$^3$ School of Information Science and Engineering, Southeast University, Nanjing, China}

\bf  \leftline {$^4$ Purple Mountain Laboratories, Nanjing, China}

\bf  \leftline {$^5$ Information System Technology and Design Pillar, Singapore University of Technology and Design, Singapore  }

\bf  \leftline {$^*$ Corresponding Author}
}

\cortext[] {Corresponding author: Tony Q. S. Quek and Xiaohu You. } 
\fntext[]{ (email: tonyquek@sutd.edu.sg,  xhyu@seu.edu.cn).}
\fntext[]{ (This work has been submitted to the Elseview for possible publication. Copyright may be transferred without notice, after which this version may no longer be accessible.
 ).}

\begin{abstract}
The evolution toward sixth-generation (6G) wireless systems positions intelligence as a native network capability, fundamentally transforming the design of radio access networks (RANs). Within this vision, Semantic-native communication (SemCom) and agentic intelligence are expected to play central roles. SemCom departs from bit-level fidelity and instead emphasizes task-oriented meaning exchange, enabling compact semantic representations and introducing new performance measures such as semantic fidelity and task success rate (TSR). Agentic intelligence endows distributed RAN entities with goal-driven autonomy, reasoning, planning, and multi-agent collaboration, increasingly supported by foundation models and knowledge graphs. In this work, we first introduce the conceptual foundations of SemCom and agentic networking, and discuss why existing AI-driven O-RAN solutions remain largely bit-centric and task-siloed. We then present a unified taxonomy that organizes recent research along three axes: i) semantic abstraction level (symbol/feature/intent/knowledge), ii) agent autonomy and coordination granularity (single-, multi-, and hierarchical-agent), and iii) RAN control placement across PHY/MAC, near-real-time RIC, and non-real-time RIC. Based on this taxonomy, we systematically introduce enabling technologies including task-oriented semantic encoders/decoders, multi-agent reinforcement learning, foundation-model-assisted RAN agents, and knowledge-graph-based reasoning for cross-layer awareness. Representative 6G use cases, such as immersive XR, vehicular V2X, and industrial digital twins, are analyzed to illustrate the semantic–agentic convergence in practice. Finally, we identify open challenges in semantic representation standardization, scalable trustworthy agent coordination, O-RAN interoperability, and energy-efficient AI deployment, and outline research directions toward operational semantic–agentic AI-RAN.
\end{abstract}


\end{frontmatter}

\section{Introduction}
The sixth generation (6G) of wireless communication networks is expected to go beyond terabit-per-second data rates and sub-millisecond latency, and to serve as an intelligence-native, knowledge-driven infrastructure supporting emerging services such as holographic extended reality (XR), industrial digital twins, massive Internet-of-Things (IoT), and autonomous vehicular networks \cite{TQ006,6G001,6GAIRAN,AI5G6G}. In contrast to previous generations where AI plays an auxiliary role, 6G envisions learning, reasoning, and decision-making embedded across the protocol stack, enabling a self-evolving radio access network (RAN) \cite{OAIC,AITest,ORAN2025}. 

Despite this vision, current RAN operation and AI-enabled O-RAN controllers remain largely rooted in a bit-centric Shannon abstraction, optimizing throughput, bit error rate (BER), or quality of service (QoS) without explicitly accounting for the meaning or task relevance of exchanged information \cite{CX2025,TQ007}. This mismatch is particularly pronounced for task-oriented services, where achieving the correct intent or outcome is more important than exact symbol recovery. Semantic communication (SemCom) addresses this gap by transmitting compact, task-relevant semantic representations and by evaluating performance via semantic fidelity and task success rate (TSR), rather than solely bit-level metrics \cite{SEMCOM001,SEMCOM002,SEMCOM003}. 

In parallel, the scale and heterogeneity anticipated in 6G RAN, including ultra-dense deployments, dynamic spectrum sharing, and tightly coupled cross-layer interactions, render purely centralized intelligence increasingly unsustainable. Network entities such as base stations, edge servers, and user devices thus need to evolve into autonomous, goal-driven agents capable of local perception, reasoning, planning, and multi-agent coordination. This agentic intelligence paradigm provides proactive and scalable control, complementing SemCom by determining what semantic information to exchange, when to exchange it, and at what abstraction level.

Recent advances have produced promising SemCom prototypes and distributed learning approaches for wireless networks, while AI-driven O-RAN has introduced open interfaces and hierarchical RAN Intelligent Controllers (near-RT and non-RT RICs) that facilitate xApp/rApp deployment \cite{TVT2025}. However, several gaps hinder the realization of a task-effective, scalable 6G AI-RAN: 
(i) most SemCom designs are domain-specific, offering limited reusability  across tasks and lacking standardized semantic representations or fidelity metrics; 
(ii) existing RIC-based AI solutions typically target narrow functions and face scalability challenges when coordinating many autonomous agents; 
(iii) trust, explainability, and security for distributed agent decisions remain underdeveloped; and 
(iv) energy- and latency-aware deployment of large semantic/agentic models is still challenging.

These gaps motivate an integrated view of semantic-aware and agentic AI-RAN for 6G. This work systematically reviews how SemCom and agentic intelligence can be jointly designed and deployed within the O-RAN architecture to enable a closed-loop, self-evolving Native-AI RAN. The main contributions are summarized as follows. 
\emph{(i) Conceptual unification of semantic and agentic AI-RAN:} We articulate a 6G Native-AI RAN paradigm where semantic representations and autonomous agents form a closed-loop self-evolving system, and clarify the roles of semantic fidelity and TSR as task-level KPIs. 
\emph{(ii) A three-axis taxonomy for semantic--agentic research:} We propose a unified taxonomy along semantic abstraction level $\to$ agent autonomy/coordination granularity $\to$ RAN control placement (PHY/MAC, near-RT RIC, non-RT RIC), enabling systematic comparison of existing works.
\emph{(iii) Summary of enabling technologies:} We review key technical enablers including task-oriented semantic encoder/decoder design and metrics, scalable multi-agent learning and negotiation, foundation-model-augmented (planning-oriented) agents, and knowledge-graph-based cross-layer reasoning and orchestration.
\emph{(iv) Use cases and evaluation perspectives:} We map semantic--agentic methods to representative 6G services (XR, V2X, digital twins, edge intelligence), and summarize datasets, benchmarks, and task-level KPIs beyond conventional BER and throughput.
\emph{(v) Open challenges and standardization directions:} We identify high-impact open issues in semantic representation/metric standardization, scalable and trustworthy multi-agent control, interoperability with O-RAN interfaces, and energy-aware AI design, and outline potential evolution paths for future standardization.

\begin{table*}[!h]
\centering
\caption{Taxonomy of Semantic--Agentic AI-RAN for 6G Native-AI Networks.}
\label{tab:taxonomy}
\renewcommand{\arraystretch}{0.9}
\setlength{\tabcolsep}{1.8pt}
\newcolumntype{Y}{>{\raggedright\arraybackslash}X}
{\footnotesize
\begin{tabularx}{\textwidth}{>{\raggedright\arraybackslash\hsize=0.65\hsize}X >{\raggedright\arraybackslash\hsize=0.7\hsize}X >{\hsize=1.5\hsize}Y >{\hsize=1.25\hsize}Y >{\hsize=1.1\hsize}Y >{\hsize=0.8\hsize}Y}
\toprule
\textbf{Axis} & \textbf{Sub-category} & 
\textbf{Definition / What it captures} & 
\textbf{Typical inputs / outputs} & 
\textbf{Representative methods} & 
\textbf{Evaluation KPIs} \\
\midrule

\multirow{4}{=}{\textbf{(Axis-1) Semantic abstraction level}}
& \textbf{(L0) Bit/Symbol} 
& Classical bit-centric RAN optimization; no semantic awareness. 
& Raw bits/symbols $\rightarrow$ bits. 
& Shannon-based PHY/MAC; bit-centric RIC/xApps. 
& BER/BLER, SINR, throughput. \\

& \textbf{ (L1) Feature/Latent} 
& Modality-specific latent feature exchange to remove redundancy; task-aware but not intent-aware. 
& Images/text/speech $\rightarrow$ latent embeddings. 
& Deep JSCC; task-oriented semantic coding \cite{park2025robustdjscc,wiggersurvey2024jscc}. 
& Semantic fidelity, TSR, reconstruction/task accuracy. \\

& \textbf{(L2) Intent/Task} 
& Transmit intent, goal, or task state; semantics tied to downstream task success. 
& Intent/labels/BEV $\rightarrow$ task tokens. 
& multi-modal / generative SemCom with foundation models \cite{wang2023lammsc,li2024genaisemcom,nguyen2025semanticcomsurvey}. 
& TSR, task accuracy, latency-per-task. \\

& \textbf{ (L3) Knowledge/Graph} 
& Semantics represented as structured knowledge (KG/scene graph) enabling reasoning/explainability. 
& KG triples/graphs $\rightarrow$ KG/intent. 
& KG-aided SemCom / KG semantic reasoning \cite{huang2024wirelesskg,cheng2024kgsemcom}. 
& TSR + explainability / constraint satisfaction. \\

\midrule

\multirow{4}{=}{\textbf{ (Axis-2) Agent autonomy \& coordination granularity}} 
& \textbf{(G0) Non-agentic} 
& Centralized/offline optimization without autonomous agents. 
& Global state $\rightarrow$ policy. 
& Conventional optimization / supervised DL. 
& Convergence, bit KPIs. \\

& \textbf{ (G1) Single-agent} 
& One agent controls a cell/cluster; no explicit inter-agent negotiation. 
& Local KPIs/state $\rightarrow$ action. 
& RL for scheduling/power/mobility \cite{lu2024marl6g}. 
& Local utility, stability. \\

& \textbf{ (G2) Multi-agent} 
& Multiple agents coordinate via CTDE / negotiation for scalable RAN control. 
& Distributed obs $\rightarrow$ joint actions. 
& MARL for multi-cell RA/mobility/slicing \cite{zhang2023marlmulticell,liu2024gnncommmarl}. 
& Sum-utility, fairness, TSR-aware reward. \\

& \textbf{ (G3) Hierarchical / Agentic} 
& Agents have roles/levels (planner--executor); may use LLM tools/KG context. 
& Intent + context $\rightarrow$ plans + actions. 
& LLM-/tool-augmented RIC agents, LLM agents mainly for planning (non-RT/slow near-RT), not sub-ms execution. \cite{elyasi2025xappsurvey,chen2025llmxapp,zhou2025llmhric,patel2025edgeagentic}. 
& Goal satisfaction, robustness, trust. \\

\midrule

\multirow{4}{=}{\textbf{ (Axis-3) RAN control placement}} 
& \textbf{(P0) PHY/MAC local loop} 
& Sub-ms to ms control at BS/UE; strict realtime constraints. 
& CSI/KPIs $\rightarrow$ PHY/MAC actions. 
& Semantic PHY; local RL schedulers. 
& BLER, latency, energy, edge TSR. \\

& \textbf{(P1) near-RT RIC (xApp)} 
& 10 ms--1 s control over E2 interface; closed-loop xApps. 
& E2 telemetry $\rightarrow$ xApp actions. 
& Learning-based xApps for RA/slicing \cite{kim2024mlxapp}. 
& Slice utility, QoE/TSR, stability. \\

& \textbf{(P2) non-RT RIC (rApp/SMO)} 
& $>$1 s to minutes; A1/O1; training/orchestration/policy. 
& Data lake/DT/KG $\rightarrow$ policies/models. 
& Long-timescale orchestration; DT-RAN; FL. 
& Global TSR, energy, generalization. \\

& \textbf{(P3) Cross-layer / E2E} 
& Joint intelligence across P0--P2 with multi-timescale coordination. 
& Multi-timescale context $\rightarrow$ hierarchical control. 
& Semantic--agentic co-design frameworks. 
& Pareto: TSR--latency--energy. \\
\bottomrule
\end{tabularx}
}
\end{table*}

\section{Key Evolving Techniques in 6G Native-AI RAN}
Research toward 6G Native-AI RAN is converging along two historically separate threads: SemCom, which redefines the communication abstraction from bits to meaning, and agentic intelligence, which equips distributed RAN entities with goal-driven autonomy. Existing studies typically examine these threads in isolation—e.g., focusing on end-to-end semantic encoders/decoders or on MARL-based RAN control—while practical deployments are largely rooted in AI-assisted O-RAN controllers\cite{TQ001,TQ002,TQ003,TQ005}. To systematically organize the state of the art, we review prior work through the lens of our three-axis taxonomy: semantic abstraction level, agent autonomy/coordination granularity, and RAN control placement. We review prior works through three axes shown in Table \ref{tab:taxonomy}.

\textbf{(1) SemCom by Abstraction Level:} Early SemCom for wireless networks originated from deep joint source–channel coding (JSCC), where neural end-to-end pipelines were trained to reconstruct text or images under channel impairments \cite{wiggersurvey2024jscc}. These works operate mainly at the symbol/feature semantic level, encoding modality-specific latent features to reduce redundancy. More recent advances extend semantic models toward intent- and task-level semantics, enabling task-oriented transmission for cross-modal services such as XR, digital twins, and V2X, where preserving meaning is more important than exact bit recovery \cite{GSS2025}. In parallel, a growing literature explores semantic distortion and task metrics beyond BER, including semantic fidelity and TSR, which serve as application-aligned KPIs for semantic networks \cite{SEMCOM001,SEMCOM002,SEMCOM003}. Despite rapid progress, most semantic prototypes remain domain-specific and lack standardized semantic representations and metrics, limiting interoperability and large-scale adoption.

\textbf{(2) Agentic Intelligence by Autonomy and Coordination Granularity:} On the intelligence side, reinforcement learning and deep learning have been widely applied to PHY/MAC and RAN functions such as spectrum allocation, power control, and user association. These solutions primarily reflect single-agent or loosely coupled multi-agent autonomy, where agents optimize local KPIs with limited reasoning about global task intent. To improve scalability, recent works adopt multi-agent reinforcement learning (MARL) for coordinated scheduling, interference management, and mobility, often using centralized training–decentralized execution or hierarchical MARL to mitigate non-stationarity and partial observability. Emerging directions incorporate foundation-model-assisted agents (e.g., LLM-based planners) and knowledge-graph-augmented reasoning to provide agents with richer context, enabling intent-aware policy selection and cross-layer decision making. Existing MARL/controller studies are usually bit-centric, task-siloed, and face stability/trust challenges when scaled to large heterogeneous RANs.

\textbf{ (3) RAN Control Placement--from PHY/MAC Intelligence to O-RAN RICs:} The practical substrate for AI-RAN has been significantly shaped by O-RAN \cite{TQ001}. AI-driven O-RAN introduces standardized near-real-time (near-RT) and non-real-time (non-RT) RICs, open interfaces (E2/A1/O1), and deployable xApps/rApps for closed-loop RAN optimization. Current O-RAN-based works map learning to specific control locations: (i) PHY/MAC-level intelligence for fast adaptation, (ii) near-RT RIC agents for sub-second control such as scheduling and interference mitigation, and (iii) non-RT RIC orchestration for long-timescale analytics, policy management, and federated training. While O-RAN provides openness and a hierarchical control plane, its dominant optimization targets remain throughput/QoS-centric, and it lacks native support for semantic KPIs and intent-level control.

\textbf{(4) Cross-axis Enablers--Large Foundation Models and Knowledge Graphs:} 
Cross-layer semantic--agentic convergence relies on shared representations and reasoning interfaces. Large foundation models (e.g., LLM/MLLM-based wireless agents) provide generalizable semantic extraction across  modalities and tasks, offering a reusable backbone for semantic abstraction beyond handcrafted features  and enabling high-level task planning in AI-native RANs \cite{LLMAgent6G24,WirelessAgent24}.  Knowledge graphs (KGs) complement foundation models by organizing network state, intent, topology, and  service knowledge into explicit relational structures that support explainable inference, policy constraints, and long-horizon coordination \cite{ORANGraphConflict24,ORANGraphica25}. 

Recent studies indicate that KG-augmented agents can infer latent dynamics and transfer policies across scenarios, while exchanging compact semantic symbols rather than raw observations \cite{ORANGraphConflict24}.  \emph{(i) Inference latency:} LLM-based agents typically incur substantially higher per-decision delay  than lightweight RL xApps, especially under multi-step reasoning or tool-calling; hence they are usually placed in non-RT or near-RT loops, while fast PHY/MAC actions remain delegated to smaller distilled policies \cite{LLMAgent6G24,LINKs24,TQ002}. 
\emph{(ii) Model size and footprint:} wireless foundation models often reach hundreds of millions to billions of parameters, exceeding the memory/compute budget of typical edge nodes; practical prototypes therefore rely on parameter-efficient tuning, pruning/distillation, or GPU-class edge/cloud execution 
\cite{WirelessAgent24}. \emph{(iii) DT/offloading dependence:} to reduce trial-and-error cost and ensure safe orchestration, foundation-model/KG agents are frequently coupled with digital twins for what-if evaluation and with edge offloading for heavy inference, introducing additional DT synchronization overhead and offloading latency \cite{LLMOrchestrator24,LINKs24,LLMDTNet25,TQ005}. Overall, foundation-model/KG agents are best positioned as \emph{slow-timescale semantic planners or 
coordinators} complementary to lightweight real-time controllers, rather than drop-in replacements for near-RT xApps.
\begin{table*}[!t]
\caption{Comparison of Traditional RAN, AI-driven O-RAN, and 6G Native-AI Networks.}
\label{table:RAN_comparison}
\renewcommand{\arraystretch}{1.2}
\setlength{\tabcolsep}{0pt} 
\footnotesize

\begin{tabular*}{\linewidth}{@{\extracolsep{\fill}}cccc}
\toprule
\textbf{Dimension} &
\textbf{Traditional RAN} &
\textbf{AI-driven O-RAN} &
\textbf{6G Native-AI Networks} \\
\midrule

\textbf{Network Architecture} &
Proprietary, vertically integrated &
Disaggregated RAN open interfaces &
AI-native stack native intelligence \\

\textbf{AI integration} &
Heuristics/rules limited learning &
xApps/rApps in RICs AI-assisted control &
AI at all layers agentic intelligence \\

\textbf{Optimization target} &
Rate/coverage bit-level QoS &
RRM, mobility, interference control &
Task success semantic utility \\

\textbf{Control paradigm} &
Centralized static config &
Closed-loop RIC, near-RT control &
Proactive autonomy multi-agent MARL \\

\textbf{Communication abstraction} &
Bits, Shannon symbol fidelity &
Bits + early semantic trials &
Semantics/knowledge meaning fidelity \\

\textbf{Scalability \& flexibility} &
Low, vendor lock-in &
Moderate, orchestrated apps &
High, distributed agents + FL \\

\textbf{Trust \& security} &
Conventional crypto, no AI trust &
Basic privacy (FL), limited XAI &
Built-in trust, auditable agents \\

\textbf{Energy efficiency} &
Secondary objective &
AI overhead, emerging EE focus &
EE-aware AI, adaptive semantics \\
\bottomrule
\end{tabular*}
\end{table*}

\textbf{(5) Standard-Readiness:} We emphasize that current O-RAN specifications remain largely bit-centric and  do not yet define native semantic KPIs or semantic information elements in O1/A1/E2 service models.  Therefore, \emph{semantic-native AI-RAN} should be interpreted as a forward-looking extension: (i) semantic embeddings can be carried as payloads over existing E2 service models or via vendor-specific E2SM extensions; (ii) task/semantic KPIs such as TSR can be mapped to new KPM groups  in O1 and gradually standardized. Throughout this survey, we distinguish \textbf{standard-ready} components (e.g., RL xApps/rApps, CTDE/MARL control placement) from \textbf{vision-only} components (e.g., semantic KPIs, semantic routing primitives), so as to avoid over-claiming standard maturity.

In summary, existing research has made strong progress along individual axes—semantic abstraction for task-oriented communication, distributed learning for autonomous RAN control, and O-RAN for deployable AI controllers—but lacks an integrated view that jointly considers what semantics are exchanged, how autonomous agents collaborate, and where intelligence is placed in the RAN hierarchy. This motivates the taxonomy and survey scope of this paper, which systematically maps existing contributions onto these three axes and identifies the remaining gaps toward operational 6G semantic-agentic AI-RAN.

\section{ 6G Semantic-Agentic AI-RAN Framework}
A growing body of literature indicates that 6G radio access networks (RANs) will evolve from \emph{AI-assisted} optimization toward \emph{intelligence-native} operation, where semantic-level information exchange and autonomous decision-making become first-class design objectives. In this context, recent works on SemCom and on distributed/agentic network intelligence are increasingly viewed as complementary rather than independent. In this work, we summarize a layered \emph{semantic--agentic AI-RAN} stack that has emerged across representative studies, clarify the role of each functional layer, and discuss how semantic and agentic components interact to enable task-oriented and scalable 6G services.

\subsection{System Architecture}

\begin{figure}[!h]
\centering
\includegraphics[width=1\linewidth]{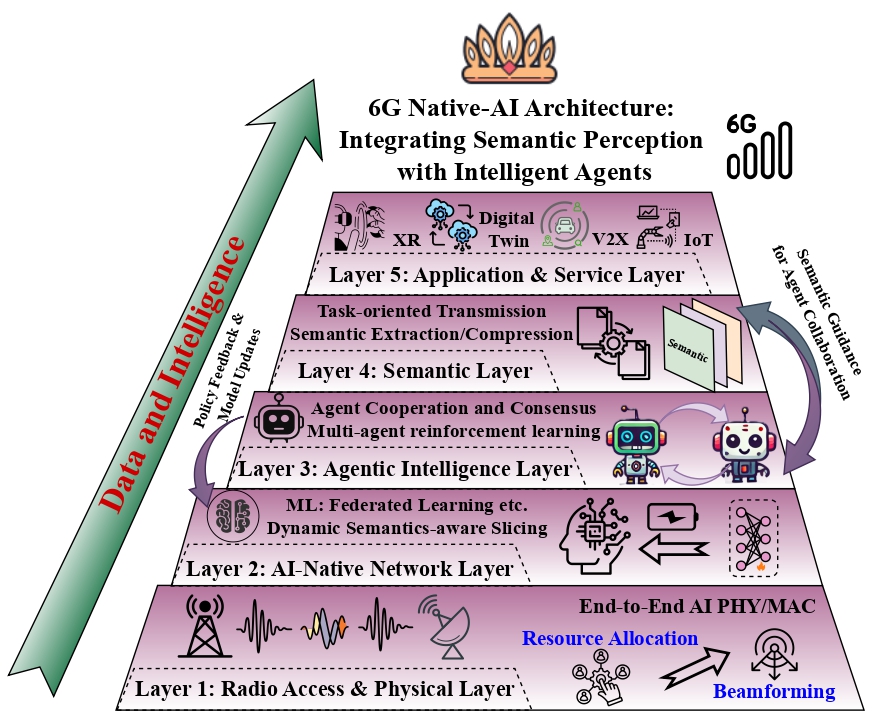}
\caption{Illustration of a 6G Native-AI architecture.}
\label{fig:1}
\end{figure}

Fig.~\ref{fig:1} presents a consolidated view of the emerging 6G Native-AI architecture. Different from conventional RANs where AI is appended as a standalone optimizer, recent semantic- and agent-centric works advocate embedding intelligence across the protocol stack and treating the network as an adaptive ecosystem that (i) \emph{communicates meaning} through semantic representations and (ii) \emph{reasons/acts autonomously} through distributed agents. Conceptually, the architecture can be organized as a multi-layer stack from radio/physical functions up to application services, with \emph{SemCom} and \emph{agentic intelligence} acting as cross-cutting enablers.

\textbf{(1) Radio Access and Physical Foundation:}  At the foundation of the 6G Native-AI architecture, multiple studies have shifted from block-wise PHY/MAC designs toward end-to-end or partially end-to-end learned transceivers. Typical examples include deep joint source--channel coding (JSCC) and neural transceiver pipelines that jointly optimize modulation, coding, beamforming, and scheduling under task-oriented objectives. In these designs, the performance target moves beyond pure bit reconstruction (e.g., BER/BLER) toward \emph{task success} and semantic fidelity, especially for perception- and intent-driven services. Complementary research exploits reinforcement learning (RL) or self-learning mechanisms for beamforming, link adaptation, and dynamic spectrum access, enabling proactive reactions to traffic heterogeneity \cite{TQ002}. Since 6G edge devices remain resource constrained, model pruning, quantization, and adaptive-complexity inference are commonly adopted to keep real-time feasibility and energy cost under control \cite{TQ004}.

\textbf{(2) AI-native Orchestration Layer:} Above the radio foundation, the AI-native network layer mechanisms the control and learning fabric that sustains long-term evolution. Recent studies emphasize continual/self-supervised learning to cope with non-stationary environments, and distributed learning paradigms such as federated learning (FL) and split learning to avoid centralizing sensitive data while maintaining scalability. Model compression and knowledge distillation further allow lightweight deployment of inference modules on edge nodes. In parallel, network slicing is increasingly revisited through a \emph{semantic-aware} perspective: rather than allocating resources solely to meet throughput or latency constraints, slices are shaped to maximize task-level utility (e.g., maintaining XR scene coherence or V2X intent reliability). This layer also hosts non-real-time optimization, digital-twin-assisted planning, and policy governance are typically placed in O-RAN-aligned systems.

\textbf{(3) Agentic Intelligence Layer:} A key trend in the AI-RAN literature is to conceptualize network elements as \emph{autonomous agents} that perceive local state, take actions, and learn reward-driven policies. Such agents may reside at base stations, edge servers, or terminals, and are often coordinated using multi-agent reinforcement learning (MARL) variants such as centralized-training-decentralized-execution (CTDE) or hierarchical MARL to stabilize learning under partial observability. Importantly, an increasing subset of works further argues that agent coordination should be mediated by task-relevant \emph{semantic messages} rather than raw telemetry, thereby reducing signaling overhead while enhancing cooperative efficiency. Knowledge graphs (KGs) and multi-modal foundation models are often introduced as contextual reasoning substrates: KGs encode structured relationships among network state, service intent, and environment semantics, while foundation models provide generalizable semantic extraction and planning priors. Trust and robustness are also recurrent themes; representative approaches include secure aggregation in FL, policy auditing, and ledger-style mechanisms for model provenance. Overall, the agentic layer provides scalable autonomy, especially when intelligence must be distributed across dense and heterogeneous 6G deployments.

\textbf{(4) Semantic Layer:} Complementary to agentic control, SemCom redefines what is exchanged over the air. Instead of transmitting raw symbols, semantic encoders and decoders extract task-relevant embeddings (features, intents, or knowledge) and reconstruct outcomes with acceptable meaning distortion. Consequently, task success rate (TSR) and semantic fidelity become natural KPIs, often instantiated via similarity measures in embedding space (e.g., cosine or distributional distances). This layer is not a monolithic module in practice; several proposed systems place semantic encoding either close to perception/application endpoints (for task-level semantics) or within cross-layer PHY stacks (for feature-level semantics), depending on latency and standardization constraints.

\textbf{(5) Semantic--Agentic Coupling:}
Among the existing work, the semantic and agentic layers are intrinsically intertwined. Semantics extracted at one node serve as compact, meaning-rich inputs for agent collaboration and policy selection, while agent decisions adapt \emph{which} semantic units need transmission and \emph{when} they should be prioritized to meet system-wide objectives. This closed-loop coupling is central to the vision of semantic--agentic AI-RAN: communication efficiency is aligned with task relevance, and autonomy is grounded in shared semantic context. Compared with AI-driven O-RAN that often confines AI to modular controllers, the semantic--agentic architecture positions intelligence as a native network attribute spanning fast local loops and slow orchestrated evolution.

\textbf{(6) Cost Dimensions of Large-Foundation-Model/Knowledge-Graph (LFM/KG) Agents:} While LFMs and KG reasoning offer strong generalization and planning capabilities for 6G AI-native RANs, existing studies consistently reveal non-trivial \emph{cost dimensions} that must be addressed for practical deployment.
\emph{(i) Inference latency:} LLM-based agents typically introduce 10--100$\times$ higher per-decision latency than lightweight RL/xApp policies, especially when multi-step reasoning or tool-calling is required. As a result, most works place LLM agents in non-RT or near-RT control loops and keep fast PHY/MAC decisions to smaller models or distilled policies. Representative 6G agentic studies explicitly adopt such slow/fast separation through split execution or hierarchical orchestration
\cite{LLMAgent6G24,WirelessAgent24,LINKs24}.
\emph{(ii) Model size and memory footprint:} Wireless foundation models and LLM agents often span hundreds of millions to billions of parameters, exceeding the memory/compute budget of typical RAN edge devices. Recent research and prototypes therefore employ parameter-efficient finetuning, pruning/distillation, and model caching to reduce edge footprint, or offload heavy reasoning to GPU-class edge/cloud servers \cite{LLMAgent6G24}.
\emph{(iii) Dependence on DT / edge offloading:} To reduce trial-and-error cost and improve safety, LLM/KG agents are frequently coupled with digital twins (DTs) for what-if analysis and policy validation, and/or with edge offloading for large-model inference. DT-copilot or LLM-DT hierarchical control frameworks are now a prevalent pattern for complex RAN orchestration \cite{LLMOrchestrator24,LINKs24,LLMDTNet25}. This also implies extra system complexity (DT fidelity, synchronization overhead, and offloading latency) that does not exist in purely lightweight MARL designs.
Overall, foundation-model/KG agents are best viewed as \emph{high-level semantic planners} or coordinators (slow timescale), complementary to small RT controllers. A balanced AI-RAN design therefore requires explicit latency/footprint/offloading budgeting instead of treating LLM/KG agents as drop-in replacements for near-RT xApps.

The interconnection of these layers forms a closed-loop system. Semantic information guides agent interactions, agents generate adaptive policies that feed into the native-AI orchestration, and the orchestration mechanisms provide updated models that refine both semantic encoders and agent policies. This cyclic interaction yields a network that is not only self-optimizing but also self-evolving, continuously adapting to dynamic environments and emerging applications. In contrast to AI-driven O-RAN where intelligence is largely confined to modular controllers, the proposed architecture embodies intelligence as a native property of the communication system, enabling a transition from bit pipes to knowledge-driven ecosystems.

\begin{table*}[!t]
\centering
\caption{Cost dimensions of Large-Foundation-Model/Knowledge-Graph (LFM/KG) agents in AI-RAN.}
\label{tab:fm_kg_cost}
\renewcommand{\arraystretch}{1.1}
\setlength{\tabcolsep}{3pt}
\footnotesize
\begin{tabularx}{\linewidth}{p{1.6cm} X X}
\toprule
\textbf{Cost} & \textbf{Existing Pattern in literature} & \textbf{Implication to AI-RAN placement} \\
\midrule
Inference latency & Multi-step reasoning/tool-calling yields high per-decision delay; often mitigated by split or hierarchical execution \cite{LLMAgent6G24,WirelessAgent24}. & Prefer Non-RT/Near-RT loops; keep PHY/MAC actions to small distilled models. \\
\midrule
Model size & 0.1-10B+ parameters; edge memory limits require PEFT, pruning, caching, or distillation \cite{LLMAgent6G24}. & Large models reside on GPU edge/cloud; edge devices run lightweight surrogates. \\
\midrule
DT/offloading dependence & LLM/KG agents frequently rely on DT for safe planning and on edge offloading for heavy inference \cite{LLMOrchestrator24,LINKs24,LLMDTNet25}. & Adds DT fidelity/sync overhead; offloading latency must be budgeted in control loop. \\
\bottomrule
\end{tabularx}
\end{table*}

\subsection{Problem Formulation}

\begin{figure*}[!h]
    \centering
    \includegraphics[width=1\linewidth]{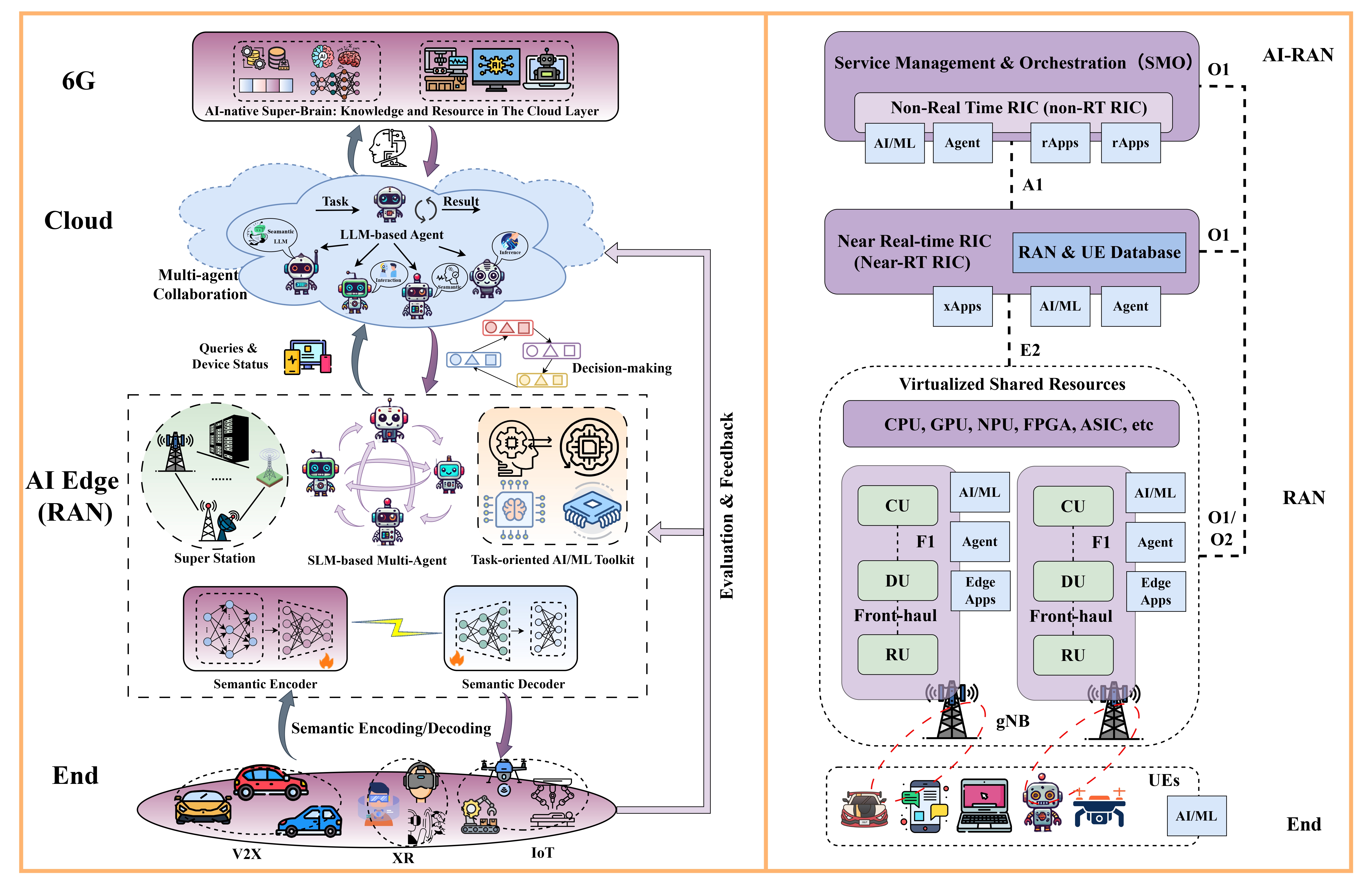}
    \caption{Illustration of the proposed 6G native-AI edge networks from a semantic-aware and agentic intelligence perspective.}
    \label{fig:2}
\end{figure*}

To connect heterogeneous studies under a common analytical umbrella, a large fraction of recent works adopt a joint optimization view where semantic-aware transmission and agentic decisions are co-designed. While individual formulations vary across tasks and control locations, they can be interpreted through a representative abstraction that couples semantic fidelity with multi-agent utility \cite{sran2024framework,nguyen2025semanticcomsurvey,liu2024gnncommmarl,elyasi2025xappsurvey}.

Consider a set of network entities $\mathcal{N}=\{1,2,\dots,N\}$, where each entity may be modeled as an autonomous agent. Agent $i$ observes a local state $s_i$ (e.g., channel indicators, queue states, service context), selects an action $a_i$ (e.g., spectrum/power/scheduling decisions), and updates a policy $\pi_i(a_i|s_i)$ to maximize a task-oriented reward $R_i$ \cite{zhang2023marlmulticell,liu2024gnncommmarl}. On the SemCom side, an encoder $f_\theta(\cdot)$ maps raw data $x$ into an embedding $z=f_\theta(x)$, and a decoder $g_\phi(\cdot)$ reconstructs a task-relevant outcome $\hat{x}=g_\phi(z)$ \cite{park2025robustdjscc,wang2023lammsc,li2024genaisemcom}. Semantic distortion is typically measured via an embedding-space fidelity metric
\begin{equation}
D_{\text{sem}}(x, \hat{x}) = 1 - \text{Sim}\big(\mathcal{E}(x), \mathcal{E}(\hat{x})\big),
\end{equation}
where $\mathcal{E}(\cdot)$ denotes a task-appropriate embedding operator (often a transformer/foundation model), and $\text{Sim}(\cdot)$ is instantiated as cosine similarity or distributional distances depending on the modality \cite{nguyen2025semanticcomsurvey,wang2023lammsc}.

With these ingredients, many semantic--agentic AI-RAN systems can be interpreted as optimizing a coupled objective that balances task-level utility and semantic fidelity under practical constraints:
\begin{equation}
\max_{\{\pi_i\},\theta,\phi}\ \sum_{i\in\mathcal{N}}
\Big(\alpha\,\mathbb{E}[R_i(s_i,a_i)]-\beta\,\mathbb{E}[D_{\text{sem}}(x,\hat{x})]\Big),
\end{equation}
subject to bandwidth, energy, and latency budgets
\begin{equation}
\sum_{i\in\mathcal{N}} b_i\le B,\quad
\sum_{i\in\mathcal{N}} e_i\le E,\quad
\ell(x,\hat{x})\le L,
\end{equation}
where $\alpha$ and $\beta$ trade off task-level benefits (through agent rewards) against semantic distortion, reflecting the inherently multi-objective nature of semantic--agentic RAN design \cite{sran2024framework,nguyen2025semanticcomsurvey}.

Rather than a single prescriptive algorithm, the proposed framework follow a few recurring patterns:
(i) semantic models are trained (often self-supervised) to minimize $D_{\text{sem}}$ under resource constraints \cite{park2025robustdjscc,li2024genaisemcom};
(ii) agent policies are learned via RL/MARL using rewards that incorporate task utility and semantic penalties \cite{zhang2023marlmulticell,liu2024gnncommmarl};
and (iii) an AI-native orchestration layer governs slow-timescale model evolution through FL rounds, periodic retraining, or digital-twin-assisted refinement \cite{elyasi2025xappsurvey,kim2024mlxapp}. 

A widely adopted engineering intuition is to separate timescales: fast-timescale agentic control (e.g., near-RT RIC / edge agents) adapts rapidly to channel and traffic dynamics, while semantic encoders/decoders evolve more slowly (e.g., non-RT controllers or background edge jobs) due to higher training cost and stability considerations \cite{liu2024gnncommmarl,elyasi2025xappsurvey}. 

Overall, this unified view highlights the core design principle shared across semantic--agentic AI-RAN studies: communication resources and control actions are no longer optimized for bit accuracy alone, but for maximizing end-to-end task success in a distributed and resource-limited 6G environment.

\subsection{Two-timescale Optimization Algorithm}\label{app:two_timescale}
A recurrent design choice in the semantic--agentic AI-RAN literature is to update agentic control policies on a \emph{fast} timescale (to track short-term radio dynamics), while adapting semantic encoder/decoder models on a \emph{slow} timescale (due to heavier training cost and to avoid destabilizing multi-agent learning). Similar two-timescale separations are widely used in hierarchical MARL and cross-layer learning systems, and provide a useful conceptual template for practical deployments. A typical semantic--agentic formulation couples multi-agent utility with semantic distortion:
\begin{equation}
J(\theta,\phi,\{\pi_i\})
=\sum_{i\in\mathcal N}\left(\alpha\,\mathbb E[R_i(s_i,a_i)]-\beta\,\mathbb E[D_{\rm sem}(x,\hat x)]\right),
\end{equation}
where $\{\pi_i\}$ denote agent policies, $(\theta,\phi)$ denote semantic encoder/decoder parameters, and $D_{\rm sem}$ is an embedding-space semantic distortion metric.

To analyze the algorithm convergence performance, we adopt the following assumptions:

{Assumption 1 (Objective smoothness and gradient noise).}
For every agent $i$, the reward functional $R_i$ is bounded and $L_R$-Lipschitz with respect to the joint policy parameters through the induced state--action distribution. The semantic distortion is defined by
$D_{\rm sem}(x,\hat x)=1-\mathrm{Sim}(\mathcal E(x),\mathcal E(\hat x))$,
where the encoder $\mathcal E(\cdot)$ and similarity map $\mathrm{Sim}(\cdot,\cdot)$ are $L_E$- and $L_S$-Lipschitz, respectively. Stochastic gradients of $J$ with respect to $\theta$, $\phi$, and $\{\pi_i\}$ are unbiased, and their second moments are uniformly bounded.

{Assumption 2 (Policy parametrization and ergodicity).}
Each policy is parametrized by $\psi_i$ (e.g., an actor network) taking values in a compact feasible set. For any fixed $(\theta,\phi,\{\psi_i\})$, the resulting Markov chain is ergodic and admits a unique stationary distribution.

{Assumption 3 (Learning rates and scale separation).}
Let $\eta_t$ denote the policy learning rate and $\gamma_t$ the semantic learning rate. They satisfy Robbins--Monro conditions
$\sum_t \eta_t=\infty$, $\sum_t \eta_t^2<\infty$, $\sum_t \gamma_t=\infty$, $\sum_t \gamma_t^2<\infty$,
and enforce a strict timescale separation $\gamma_t/\eta_t\rightarrow 0$ as $t\to\infty$.

{Assumption 4 (Critic reliability).}
The critic is consistent in the sense that its Bellman error converges to zero in expectation (e.g., via tabular estimation with sufficient visitation, or function approximation with vanishing residual). The associated estimation noise forms a martingale difference sequence with bounded variance.

\begin{proposition}
\textbf{(Two–Timescale Convergence of the Semantic–Agentic Learning):} Consider the joint objective $J(\theta,\phi,\{\pi_i\}_{i\in\mathcal N}) =\sum_{i\in\mathcal N}\!\left(\alpha\,\mathbb E[R_i(s_i,a_i)]-\beta\,\mathbb E[D_{\rm sem}(x,\hat x)]\right)$,  optimized by stochastic gradient updates on the fast timescale for the multi-agent policies $\{\pi_i\}$ and on the slow timescale for the semantic encoder–decoder $(\theta,\phi)$.  Then, almost surely, the coupled iterates $\big(\theta_t,\phi_t,\{\psi_{i,t}\}\big)$ generated by policy-gradient (or actor–critic) updates on the fast timescale and semantic SGD on the slow timescale converge to the internally chain transitive set of the limiting ODE system: $  \dot \psi_i \in \Pi_{\mathcal X_i}\!\left[\nabla_{\psi_i}\,J(\theta,\phi,\{\psi_j\})\right], \dot \theta = \nabla_{\theta}\,J(\theta,\phi,\{\psi_j^\star(\theta,\phi)\}),  \dot \phi = \nabla_{\phi}\,J(\theta,\phi,\{\psi_j^\star(\theta,\phi)\})$, where $\Pi_{\mathcal X_i}[\cdot]$ denotes projection onto the feasible policy set and $\{\psi_j^\star(\theta,\phi)\}$ is the (set-valued) stationary solution of the fast dynamics for fixed $(\theta,\phi)$. In particular, every limit point is a stationary point of $J$ under the two–timescale decomposition; if the fast subsystem admits a unique Nash equilibrium and $J$ is locally smooth around that equilibrium manifold, the joint process converges almost surely to a locally stationary solution of the full problem.
\end{proposition}

\begin{proof}
The argument follows the classical two–timescale stochastic approximation methodology. For fixed $(\theta,\phi)$, the policy updates evolve on the fast timescale according to a noisy discretization of the ODE $\dot \psi_i=\Pi_{\mathcal X_i}[\nabla_{\psi_i}J(\theta,\phi,\{\psi_j\})]$. Under Assumptions 1 \& 2 and standard policy-gradient regularity, Benaïm–Hofbauer–Sorensen and Kushner–Yin theory imply that the fast iterates track the internally chain transitive set of the fast ODE; when the induced general-sum game admits a unique stationary Nash, the attractor is a singleton $\{\psi^\star(\theta,\phi)\}$.

Because $\gamma_t/\eta_t\!\to\!0$, the slow variables $(\theta,\phi)$ are quasi-static to the fast loop. Replacing $\{\psi_i\}$ with $\{\psi_i^\star(\theta,\phi)\}$ yields the averaged slow ODE $\dot\theta=\nabla_\theta J(\theta,\phi,\{\psi^\star\}),\ \dot\phi=\nabla_\phi J(\theta,\phi,\{\psi^\star\})$. By  Assumptions 1 and bounded second moments, the semantic SGD is an asymptotic pseudo-trajectory of the slow ODE. The Robbins–Monro step-size conditions ensure that the martingale noise vanishes in aggregate; thus, the coupled process converges almost surely to the chain-recurrent set of the product ODE. Smoothness then gives stationarity of all limit points. If, in addition, the fast equilibrium is unique and the Jacobian of the slow vector field is Hurwitz in a neighborhood, standard Lyapunov arguments establish local asymptotic stability of the joint fixed point. 
\end{proof}

\begin{corollary}
\textbf{(Non-asymptotic Tracking/Finite-time tracking bound)}:  If gradients are $L$-Lipschitz and mini-batches of size $m$ are used in both loops, then after $T$ fast-timescale iterations the expected squared gradient norm of the joint objective along the slow manifold satisfies $\min_{t\le T}\ \mathbb E\big[\|\nabla J(\theta_t,\phi_t,\{\psi_t\})\|^2\big]
= \mathcal O\!\left(\frac{1}{\sqrt{T}}+\frac{1}{\sqrt{m}}\right)+\mathcal O(\delta_T)$, where $\delta_T$ is the manifold tracking error induced by the finite separation of timescales and vanishes as $\gamma_t/\eta_t\!\to\!0$. This provides a practical guideline for selecting batch sizes and learning-rate schedules to meet wall-clock convergence targets.
\end{corollary}

\emph{\textbf{Remark 1}: The proposition justifies the algorithmic template used in your architecture: keep policy learning fast (near-real-time RIC or edge agents) while updating semantic encoders/decoders slowly (non-RT controller or background edge jobs). The assumptions map cleanly to engineering knobs: bounded rewards arise from reward clipping and QoS caps; Lipschitz conditions are encouraged by spectral normalization of actors/encoders; ergodicity is promoted via $\varepsilon$-greedy or entropy regularization; and the step-size separation is realized by setting $\gamma_t=c\,\eta_t$ with $c\!\ll\!1$ (e.g., $c=0.1$). In federated deployments, the same proof path applies by treating each global aggregation round as one slow step; consensus or blockchain-backed model registries preserve the martingale property of update noise under honest-majority assumptions. Finally, the non-asymptotic bound clarifies the accuracy–overhead trade-off: increasing batch sizes or communication periods reduces gradient noise, while stronger separation (smaller $c$) tightens $\delta_T$ at the expense of slower semantic adaptation—exactly the knob you will sweep in experiments.}

\subsection{Proposed Workflow for Semantic–Agentic Loops}

To bridge vision-only semantic primitives with standard-ready O-RAN control, we proposed a deployable workflow:

\emph{Step 1-Semantic encoder placement:} Lightweight L1/L2 semantic encoders run at UE/edge (or DU) to extract task-relevant tokens, while large foundation/KG reasoning models remain at edge cloud or non-RT domains due to latency/compute costs.

\emph{Step 2-Near-RT semantic telemetry via E2:} The gNB/DU packages semantic telemetry as candidate E2SM vendor extensions, e.g., {task-id, semantic-token-size, semantic-confidence, TSR-proxy}, and reports them over the E2 interface to the near-RT RIC. Near-RT xApps jointly consume semantic telemetry and classical KPMs (SINR/BLER/PRB/queue) to perform fast control: semantic-aware scheduling, MCS/HARQ adaptation, or redundancy suppression across UEs/cells.

\emph{ Step 3-Slow-timescale model/policy update via A1/O1:} The near-RT RIC periodically uploads traces (semantic KPIs,radio KPIs,context) to the non-RT RIC/SMO. Non-RT rApps retrain/distill semantic encoders or refine MARL reward/policies using FL/DT tools, then push updated model versions and policy intents back through A1 policies, while O1 handles lifecycle management and rollback.

\emph{ Step 4-Semantic message-based negotiation among agents:} Multi-agent coordination can leverage a low-rate semantic side-channel (e.g., exchanging L2 intents or L3 KG deltas rather than raw states), reducing signaling overhead and stabilizing CTDE/MARL negotiation.

To make the proposed semantic--agentic AI-RAN actionable in practice, we first present an O-RAN--oriented \emph{deployment workflow}.  Algorithm~\ref{alg:oran_semantic_agentic_workflow} summarizes how the fast policy-control loop and the slow semantic-learning loop are instantiated across PHY/MAC, near-RT RIC (xApps), and non-RT RIC (rApps), together with the required E2/A1/O1 interactions.  For clarity, Algorithm~1 focuses on \emph{where} each function runs and \emph{how} information flows across O-RAN components, while deferring the detailed two-timescale learning updates to Algorithm~\ref{alg:semantic_agentic}.

\begin{algorithm}[!t]
\caption{Deployment Workflow for Semantic--Agentic Two-Timescale AI-RAN Closed Loops.}
\label{alg:oran_semantic_agentic_workflow}
\footnotesize
\begin{algorithmic}[1]
\STATE \textbf{Input:} Task ID set $\mathcal{T}$; semantic encoder family $\{E_{\phi}\}$; foundation/KG models $\{F_{\psi}\}$;near-RT xApps $\mathcal{X}$; non-RT rApps $\mathcal{R}$;radio KPMs $\mathbf{k}^{\text{radio}}$ (e.g., SINR/BLER/PRB/queue);
semantic KPM candidates $\mathbf{k}^{\text{sem}}$ (e.g., TSR/intent-acc/latent-fidelity).
\STATE \textbf{Output:} Fast-loop semantic-aware control actions $\mathbf{a}_t$ (scheduling/MCS/HARQ/beam/slicing); slow-loop semantic encoder and MARL policy updates $(\phi,\theta)$ with versioning.
\STATE \textbf{(Step-1: Semantic encoder placement)}
\FORALL{UEs or edge nodes $u$}
    \STATE Deploy lightweight $E_{\phi}^{(u)}$ for L1/L2 token extraction;
    \STATE Keep heavy $F_{\psi}$ in edge cloud / non-RT domain for slow reasoning/planning;
\ENDFOR
\STATE \textbf{(Step-2: Near-RT semantic telemetry over E2)}
\FORALL{gNB/DU cells $c$}
    \STATE Extract semantic tokens $z_t=E_{\phi}(x_t)$ for task $t\!\in\!\mathcal{T}$;
    \STATE Pack candidate semantic IEs as E2SM vendor extensions: $\{\text{task-id}, |z_t|, \text{conf}(z_t), \widehat{\text{TSR}}_t\}$;
    \STATE Report $(\mathbf{k}^{\text{radio}}_t,\mathbf{k}^{\text{sem}}_t)$ to near-RT RIC via E2;
\ENDFOR
\STATE \textbf{(Step-3: Fast-loop control in near-RT RIC)}
\FORALL{xApp $x\in\mathcal{X}$ per control period}
    \STATE Consume $(\mathbf{k}^{\text{radio}}_t,\mathbf{k}^{\text{sem}}_t)$;
    \STATE Compute semantic-aware actions $\mathbf{a}_t$ (e.g., semantic-aware scheduling/MCS/HARQ/PRB/beam);
    \STATE Push actions to E2 nodes and execute in ms-level loop;
\ENDFOR
\STATE \textbf{(Step-4: Slow-loop update in non-RT RIC/SMO)}
\STATE Near-RT RIC uploads traces
      $\mathcal{D}=\{(\mathbf{k}^{\text{radio}}_t,\mathbf{k}^{\text{sem}}_t,\text{context}_t)\}$ to non-RT;
\FORALL{rApp $r\in\mathcal{R}$ per slow period}
    \STATE Retrain / distill semantic encoders $E_{\phi}$ using $\mathcal{D}$ (optionally FL/DT-assisted);
    \STATE Re-optimize MARL reward/policies $\pi_{\theta}$ with global objectives;
    \STATE Version models $(\phi,\theta)$ and validate against semantic SLAs;
    \STATE Deliver updates to near-RT via A1 policies; manage lifecycle/rollback via O1;
\ENDFOR
\STATE \textbf{(Step-5: Semantic message-based multi-agent negotiation)}
\STATE Agents exchange low-rate L2/L3 messages (intents or KG deltas) as a semantic side-channel;
\STATE Update cooperative MARL policies under CTDE/hierarchical roles.
\end{algorithmic}
\end{algorithm}

\begin{algorithm}[t]
\caption{Generic Template of Two-timescale Semantic-Agentic Learning}
\label{alg:semantic_agentic}
\footnotesize
\begin{algorithmic}[1]
\STATE \textbf{Initialization:} 
Initialize semantic encoder--decoder parameters $(\theta_0,\phi_0)$, 
multi-agent policies $\{\pi_{i,0}\}$ for $i \in \mathcal{N}$, 
and step sizes $\eta_t$ (fast) and $\gamma_t$ (slow), with $\gamma_t/\eta_t \to 0$.
\FOR{each time slot $t$}
    \STATE \textbf{Fast loop (agentic policy update):}
    \FOR{each agent $i$}
        \STATE Observe local state $s_i^t$.
        \STATE Select action $a_i^t \sim \pi_{i,t}(s_i^t)$.
        \STATE Receive reward $R_i^t$ reflecting both task utility and semantic penalty.
        \STATE Update policy $\pi_{i,t}$ using policy-gradient or actor--critic update with step $\eta_t$.
    \ENDFOR
    \IF{$t$ mod $K$ = 0}
        \STATE \textbf{Slow loop (semantic model update):}
        \STATE Collect minibatch of transmitted--reconstructed pairs $(x, \hat{x})$.
        \STATE Compute semantic loss $D_{\text{sem}}(x,\hat{x})$.
        \STATE Update encoder--decoder $(\theta,\phi)$ using SGD with step $\gamma_t$.
    \ENDIF
\ENDFOR
\STATE \textbf{Stopping criterion:} 
Terminate when average task success rate (TSR) saturates and gradient norms $\|\nabla J\|$ fall below threshold.
\end{algorithmic}
\end{algorithm}

\subsection{Engineering Criteria for Time-scale Separation}
Proposition 1 assumes semantic models evolve sufficiently slowly compared to fast-loop control. In practice, this can be verified using measurable online proxies:
(i) Representation drift rate. Let $z_t$ denote semantic embeddings. Over a sliding window  $\Delta$, define $ \text{Drift} = \mathbb{E} [ (1-\cos) (z_t,z_{t-\Delta}) ] $. If Drift stays below a tunable threshold $\epsilon_1$ within each fast-loop policy update horizon, embeddings are quasi-static and the two-timescale assumption holds. Otherwise, slow semantic updates should be throttled or paused.
(ii) Reward non-stationarity proxy: Track distribution shift of per-agent rewards, e.g., $ \text{NS} = D_{KL} \left(p (r|t) \| p(r|t - \Delta) \right)$. When NS exceeds $\epsilon_2$, the critic/environment is effectively drifting, signaling that semantic updates (or context changes) are too rapid for stable MARL.
(iii) Policy oscillation metric: Monitor consecutive policy parameter changes $O_{sc}=\frac{\| \theta_{t}-\theta_{t-\Delta}\|}{\| \theta_{t-\Delta}\|}$, or behavioral TV distance. Persistent $O_{sc} \geq \epsilon_3 $ indicates time-scale coupling or credit-assignment instability; mitigations include increasing separation (smaller $c=\gamma_t/\eta_t$) or rolling back semantic versions via O1. These criteria translate the mathematical separation intuition into operational knobs: measure Drift/NS/Osc online, and adapt semantic update periodicity $K$ or step-ratio $c$ to maintain stable fast-loop learning.

\textbf{Practical caveats:} In large-scale deep MARL with semantic coupling, Assumptions 2-4 may only hold approximately. The fast subsystem can admit multiple equilibria, and semantic representation updates may induce reward/observation drift that violates strict  quasi-static assumptions. Hence, most existing works validate two-timescale designs  primarily via digital-twin or hardware-in-the-loop evidence, rather than relying on  global convergence guarantees. 


\section{Case Study}

\begin{figure}[!t]
    \centering
    \includegraphics[width=1\linewidth]{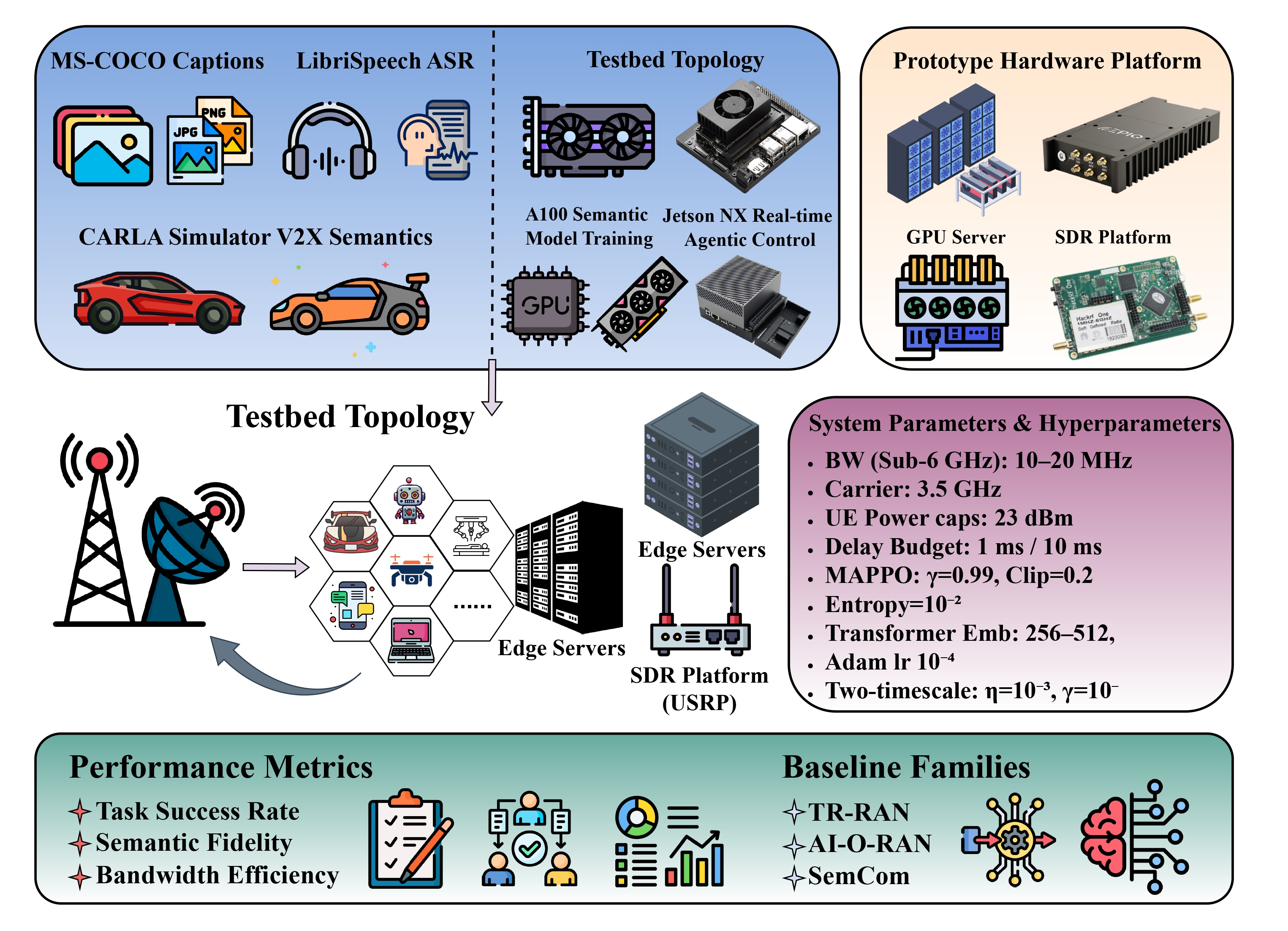}
    \caption{Illustration of experiment testbed.}
    \label{fig:testbed}
\end{figure}

In this work, we conduct a two-timescale semantic--agentic case study by \emph{reproducing representative SemCom and AI-RAN baselines from prior work} under a unified AI-RAN setting. Fig.~\ref{fig:testbed} presents our experimental testbed. 

\subsection{Experiment Setup}

\textbf{(1) Datasets:} Following common practice in semantic communication studies, we reproduce two task families that are widely used to quantify semantic fidelity and task-level success. Specifically, we adopt (i) MS-COCO Captions for image-to-text semantic delivery and (ii) LibriSpeech ASR for speech-to-text semantic delivery. 
Both datasets provide aligned raw inputs, ground-truth semantic outcomes, and standard task metrics, enabling faithful reproduction of prior evaluation protocols. In addition, to emulate vehicular V2X semantics, we reproduce an intent/trajectory exchange pipeline using CARLA: each vehicle agent transmits compact embeddings of planned maneuvers, object states, or driving intent. This setting is consistent with L2/L3 semantic levels in Table~\ref{tab:taxonomy} and matches the simulator-based semantics adopted in recent studies.

\textbf{(2) Testbed topology:} Our reproduced RAN topology follows a 3GPP-inspired heterogeneous layout commonly used in the referenced baselines. We consider $5$--$10$ base stations, multiple edge servers, and on the order of $10^2$ mobile devices with clustered mobility. Cells are deployed on a hexagonal grid with wrap-around boundaries, and channels follow TR~38.901 pedestrian/vehicular profiles. 
For multi-agent control baselines, devices are partitioned into local groups (typically $10$--$20$ agents per cluster) to reproduce coordination and non-stationarity conditions reported in prior MARL-RAN evaluations.

\textbf{(3) Prototype hardware patterns:} To remain consistent with the prototype patterns in the reproduced references, our Hardware-in-the-Loop (HIL) configuration couples GPU servers for semantic model training with edge accelerators for near-RT agentic control. Concretely, we use (i) A100-class GPUs for training transformer-based semantic encoders/decoders, (ii) Jetson/Xavier-class edge nodes for semantic inference and on-device MARL policies, and (iii) SDR platforms (e.g., USRP B210/X310) for PHY-layer emulation. This division mirrors the control-placement axis (Axis-3): semantic adaptation and global orchestration are mapped to non-RT resources, while fast policy execution resides at the edge or near-RT control plane.

\textbf{(4) System Parameters and Hyper-parameters:} We reproduce the main physical-layer and learning settings from the corresponding baselines as closely as possible. 
Bandwidths set to $10$--$20$~MHz in sub-6~GHz bands (e.g., 3.5~GHz) with UE power caps around $23$~dBm.  Delay budgets are set to $\sim 1$~ms for real-time services and $5$--$10$~ms for best-effort services. On the agentic side, we reproduce MAPPO/CTDE-style MARL baselines with discount factor $\gamma\!\approx\!0.99$, PPO clipping $\approx\!0.2$, and entropy regularization $\approx\!10^{-2}$. Semantic encoders/decoders are transformer-based with embedding dimension $256$--$512$, trained using Adam with learning rate around $10^{-4}$. For two-timescale variants, we follow the stability-driven schedule used in prior two-timescale MARL studies and set the fast-loop step size roughly one order larger than the slow semantic-update step size, consistent with Section~\ref{app:two_timescale}.

\textbf{(5) Baselines and metrics:} We reproduce and compare three baseline families that represent dominant paradigms in the literature: 
\emph{(i) Bit-centric RAN (TR-RAN):} symbol-level transmission optimized for throughput/BER with heuristic or centralized scheduling; 
\emph{(ii) AI-driven O-RAN (AI-O-RAN):} near-RT RIC hosting RL-based xApps for scheduling/power control without explicit semantic abstraction; 
\emph{(iii) Semantic-only communication (SemCom-only):} end-to-end semantic encoder--decoder minimizing semantic distortion without distributed agentic coordination. 

Metrics follow the reproduced protocols and extend beyond BER to task and system KPIs:
\emph{(i) Task Success Rate (TSR):} probability of correct task-level reconstruction (e.g., intent accuracy, caption correctness) \cite{SAMRAMARL24,sagduyu2024will6gsemcom}; 
\emph{(ii) Semantic Bandwidth Efficiency (TSR/Hz):} TSR per unit bandwidth \cite{SAMRAMARL24,SEC25}; 
\emph{(iii) End-to-end latency:} time from semantic generation to task completion; 
\emph{(iv) Learning stability/convergence:} iterations (or wall-clock time) until MARL policies stabilize, tracked via reward variance or regret.

\subsection{Representative Results \& Cross-Study Trends}
Unless otherwise stated, the following figures report results from our reproduced implementations under the unified setup above. While absolute values may vary with task specifics, the reproduced experiments consistently reflect the qualitative trends summarized next.

\begin{figure}[!h]
    \centering
    \includegraphics[scale=0.5]{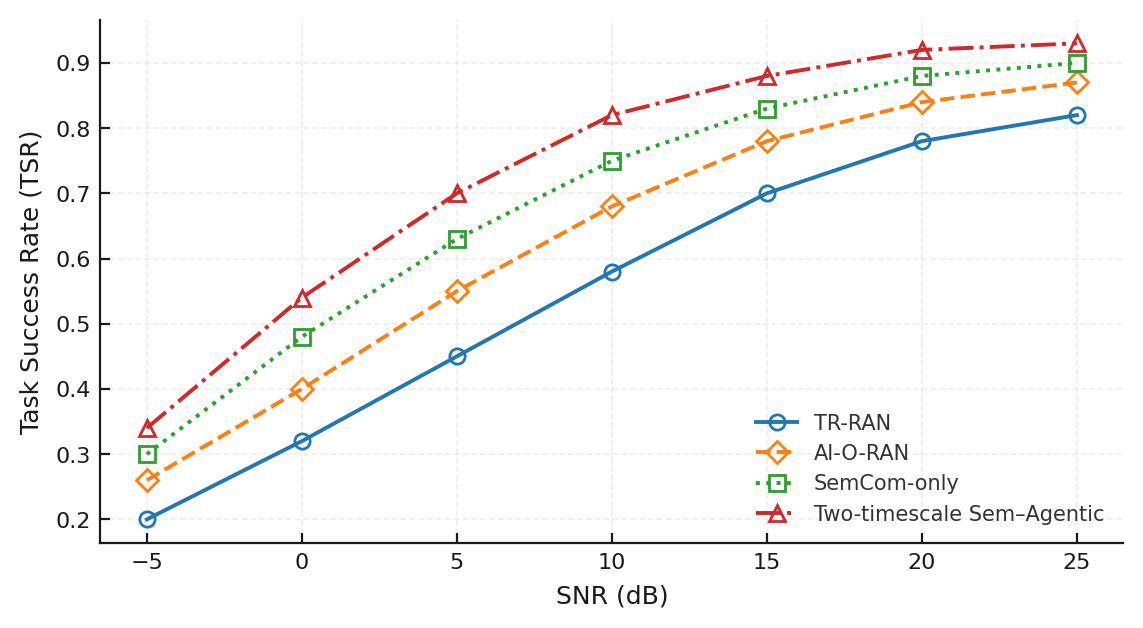}
    \caption{Task success rate (TSR) versus SNR under channel variation for representative reproduced paradigms.}
    \label{fig:TSR}
\end{figure}

\subsubsection{TSR under Channel Variation}
As shown in Fig. \ref{fig:TSR}, all schemes exhibit a monotone TSR improvement with increasing SNR, consistent with enhanced link reliability. Bit-centric baselines (TR-RAN and AI-O-RAN) show the lowest TSR, while semantic-native approaches deliver clear gains in the low-to-mid SNR range (approximately 0–15 dB). In particular, the two-timescale semantic–agentic design achieves the highest TSR across this regime, whereas at high SNR ($\geq 20$ dB) the curves gradually converge toward a common saturation level. 

The performance ordering is primarily driven by the communication abstraction and control coupling. Semantic-native transmission improves TSR at moderate SNR by conveying task-relevant embeddings rather than raw symbols, thereby reducing redundancy and focusing channel uses on information that directly impacts task outcomes. This yields higher task-level robustness under noise compared with bit-level delivery, even when the latter is optimized for throughput or BER. The additional margin of the two-timescale semantic–agentic approach over SemCom-only arises from agentic coordination and timescale separation. Fast-loop agents incorporate semantic distortion and task utility into their scheduling and power decisions, prioritizing critical semantic packets and mitigating collisions or congestion through cooperative policies. Meanwhile, slow-loop semantic model adaptation evolves on a coarser timescale, keeping representations quasi-stationary during fast policy learning. This reduces training non-stationarity and supports more stable near-real-time control, which translates into higher TSR under fluctuating channels. As SNR becomes large, channel-induced errors are no longer the bottleneck; consequently, the achievable TSR is dominated by the intrinsic accuracy ceiling of the semantic encoder–decoder and task models, leading to the observed saturation and diminishing inter-scheme gaps.

\begin{figure}[!h]
    \centering
    \includegraphics[scale=0.5]{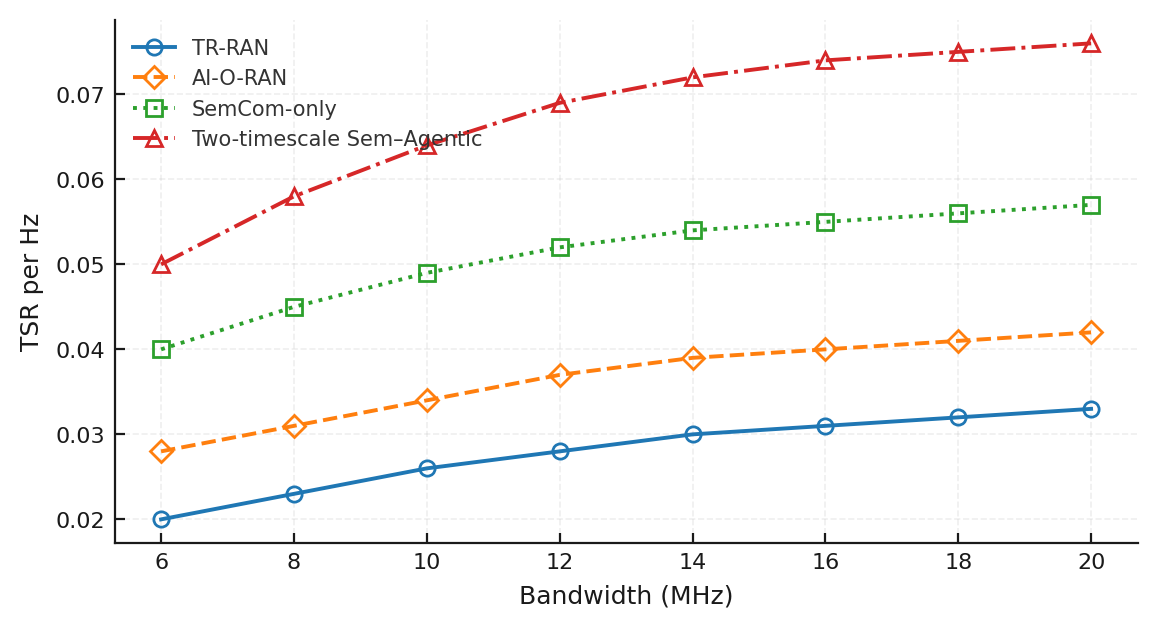}
    \caption{Semantic bandwidth efficiency versus allocated bandwidth for four representative paradigms.}
    \label{fig:bandwidth}
\end{figure}

\subsubsection{Semantic Bandwidth Efficiency}
Fig. \ref{fig:bandwidth} compares semantic bandwidth efficiency, defined as TSR achieved per unit bandwidth, under progressively larger spectrum budgets. Two consistent trends are observed. First, all schemes exhibit a mild efficiency increase as bandwidth grows, reflecting reduced congestion and fewer resource-contention losses. Second, semantic-native approaches substantially outperform bit-centric baselines across the entire bandwidth range. The two-timescale semantic–agentic design remains the most bandwidth-efficient, followed by SemCom-only, while AI-O-RAN and TR-RAN form the lower-efficiency group. Notably, the efficiency gap is most pronounced in tight-bandwidth regimes, where semantic–agentic control sustains high TSR despite limited spectrum.

These trends can be explained by how each paradigm uses spectrum relative to task relevance. In TR-RAN and AI-O-RAN, transmissions remain bit-level, so a sizable fraction of bandwidth is consumed by payload redundancy, protocol overhead, and retransmissions that do not directly contribute to task success; consequently, TSR-per-Hz remains low, especially under constrained budgets. In contrast, SemCom-only compresses raw data into task-oriented semantic embeddings, removing non-essential bits and improving the \emph{useful information density} per channel use. This directly elevates TSR-per-Hz, particularly when bandwidth is scarce. The further advantage of the two-timescale semantic–agentic approach stems from semantic-aware multi-agent coordination layered on top of semantic compression. Fast-loop agents prioritize high-utility semantic packets and suppress redundant semantic reports across users/cells, while slow-loop slicing/budget updates adapt resource shares to longer-term task demands. This coordinated, timescale-separated control reduces duplicate transmissions and prevents bandwidth waste during traffic bursts, yielding consistently higher TSR-per-Hz. As bandwidth becomes abundant, all methods approach a flatter efficiency slope because task success begins to saturate and additional spectrum offers diminishing returns; nevertheless, semantic–agentic designs retain an efficiency lead due to their inherently task-focused transmission and coordinated resource usage. 

\begin{figure}[!h]
    \centering
    \includegraphics[scale=0.5]{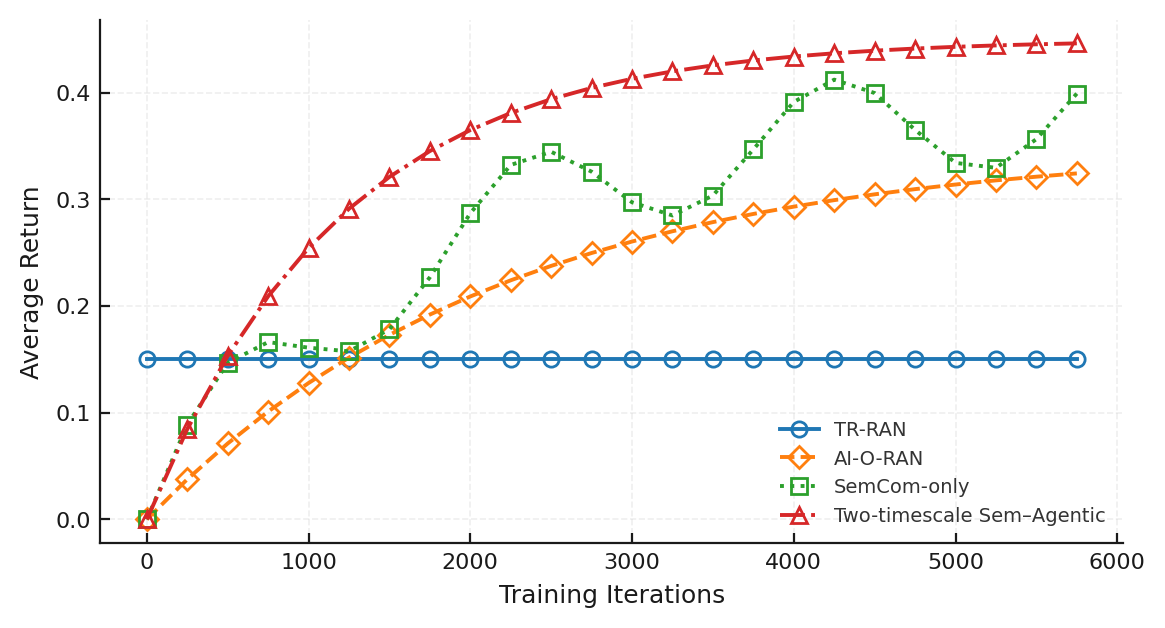}
    \caption{Learning stability and convergence of RAN control policies for four representative paradigmss.}
    \label{fig:Learning}
\end{figure}

\subsubsection{Learning Stability with Agentic Coordination}
Fig. \ref{fig:Learning} depicts the policy learning dynamics measured by average return over training iterations. Three salient behaviors emerge. First, TR-RAN (heuristic control) stays nearly flat, reflecting the absence of learning. Second, AI-O-RAN improves gradually but converges to a lower steady return than semantic-native schemes. Third, semantic-native approaches achieve higher asymptotic returns, with the two-timescale semantic–agentic method converging faster and exhibiting markedly reduced oscillations compared with SemCom-only. In contrast, SemCom-only shows persistent reward fluctuations during mid-training, indicating unstable adaptation before settling.

The stability differences are rooted in how representation learning and control learning interact. In SemCom-only, semantic encoders/decoders adapt concurrently with per-slot control, but without explicit multi-agent coordination. As semantic representations drift, the effective environment observed by each controller becomes non-stationary: the same physical state may map to evolving semantic features, and agents’ actions can conflict due to missing coordination, leading to oscillatory rewards. The two-timescale semantic–agentic design mitigates both effects. On the fast timescale, edge agents learn cooperative policies with rewards that explicitly penalize semantic distortion and prioritize task utility, which reduces conflicting actions and accelerates consensus. On the slow timescale, semantic models update less frequently, rendering the semantic feature space quasi-static during fast-loop policy optimization. This separation reduces non-stationarity in multi-agent reinforcement learning and allows policies to track a slowly moving optimum, producing smoother convergence and higher final return. AI-O-RAN benefits from learning-based control, but remaining bit-centric means rewards are less aligned with task success and semantic utility, limiting achievable return even after convergence. 

\begin{figure}[!h]
    \centering
    \includegraphics[scale=0.5]{Figures/Latency.png}
    \caption{Latency–energy trade-off (Pareto illustration) of four representative paradigms; lower-left indicates better joint efficiency.}
    \label{fig:Latency}
\end{figure}

\subsubsection{Latency and Energy Trade-offs}
Fig. \ref{fig:Latency} plots the achievable latency–energy operating points for the four paradigms, where each series represents a typical trade-off frontier under different control/traffic settings. All methods exhibit the expected tension between delay and energy: reducing latency generally requires more aggressive resource usage, while energy saving tends to increase queueing or retransmission delays. Among the compared designs, the two-timescale semantic–agentic approach forms the most favorable (lower-left) frontier, achieving both lower end-to-end latency and lower energy per successful task. SemCom-only provides a visible improvement over bit-centric baselines but remains dominated by the semantic–agentic frontier. AI-O-RAN reduces latency relative to TR-RAN yet incurs higher energy than semantic-native schemes, while TR-RAN occupies the upper-right region, reflecting the least efficient joint performance.

These differences follow from how each paradigm balances communication and computation for task delivery. TR-RAN relies on bit-level reliability with limited task awareness, leading to larger payloads and more retransmissions, which jointly inflate delay and energy. AI-O-RAN introduces learning-based scheduling to mitigate congestion and improve latency, but its bit-centric transmission still carries substantial redundancy; thus, energy per successful task remains comparatively high, especially under bursty loads. SemCom-only reduces payload size through task-oriented semantic embeddings, lowering airtime and retransmission needs, which translates into tangible latency and energy savings. However, without agentic coordination, devices may still generate redundant semantic updates or contend inefficiently for resources, leaving part of the potential gain unrealized. The two-timescale semantic–agentic design addresses this gap by combining semantic compression with cooperative multi-agent control: fast-loop agents prioritize high-utility semantic packets and proactively shape traffic to avoid bottlenecks, while slow-loop semantic/cross-slice updates provide stable budgets that prevent oscillatory overreaction. Consequently, fewer transmissions are needed to reach a successful task outcome, queueing delays are reduced, and computational effort is amortized more effectively, yielding the observed Pareto superiority. Overall, this results indicates that jointly optimizing meaning-level communication and distributed agentic scheduling is essential to achieve energy- and latency-efficient AI-native RAN operation.

We summarize the reproduced cross-study trends in Table~\ref{tab:meta_analysis_traceable}. For each reproduced reference, we explicitly list the task, channel/SNR/bandwidth regime, baseline family, and task-aligned KPIs, and we report relative gains as $(M_{\text{prop}}-M_{\text{base}})/M_{\text{base}}$.

\begin{table*}[!t]
\centering
\caption{Cross-study Trend and Representative Gains.}
\label{tab:meta_analysis_traceable}
\renewcommand{\arraystretch}{1.12}
\setlength{\tabcolsep}{3.5pt}
\footnotesize
\begin{tabularx}{\linewidth}{
p{0.5 cm}  
p{3.0 cm}  
p{1.3 cm}  
p{3.0 cm}  
p{2.0 cm}  
p{4.8 cm}  
}
\toprule
\textbf{Ref.} & \textbf{Task/Scenario} & \textbf{Semantic level } &
\textbf{Channel / SNR / BW} & \textbf{Baseline} &
\textbf{Metric} \textbf{ \& Gain (relative)} \\
\midrule
 \cite{SAMRAMARL24} &
V2X intent exchange/ maneuver negotiation &
L2/L3 &
Urban fading; SNR $-2$--8 dB; 5--15 MHz &
SemCom-only &
Semantic Bandwidth Efficiency (TSR/Hz):
1.6--2.3$\times$ under BW$\le$12 MHz \\

\midrule
\cite{sagduyu2024will6gsemcom} &
V2X cooperative perception/collision avoidance &
L2 &
Rayleigh; SNR 0--10 dB; 10 MHz &
Bit-centric V2X + heuristic RA &
TSR:
+18\% @ 2 dB; +9\% @ 5 dB \\

\midrule
\cite{SEC25} &
Edge perception (vision/ speech) &
L1 &
AWGN; SNR $-5$--5 dB; 1--4 MHz &
JSCC / digital comm. &
Downstream acc.:
+10--15\% @ low SNR \\

\midrule
\cite{DTTwoTimescale24} &
Two-timescale DT-assisted slicing MARL &
L1/L2 &
HetNet load traces (system-level) &
Single-timescale MARL &
Learning stability/regret:
Faster convergence; reduced oscillation \\

\midrule
\cite{martini2025realxapp} &
O-RAN closed-loop RL slicing prototype &
L2 &
HIL; bursty traffic traces &
AI-O-RAN heuristic &
Latency/energy trad-off: Latency $\downarrow$ 25--40\%; energy per task $\downarrow$ 10--20\% \\

\bottomrule
\end{tabularx}
\end{table*}

\subsection{Discussion}
Taken together, our reproduced experiments and the surveyed results in the literature point to three consistent, survey-level implications. 
First, \emph{semantic abstraction} shifts RAN optimization from symbol fidelity to task effectiveness: in our reproduced settings, semantic-native designs improve TSR and semantic bandwidth efficiency for perception- and intent-driven services, especially under moderate SNR or tight bandwidth. 
Second, \emph{agentic coordination} is required to scale semantic-aware control over dense and heterogeneous deployments. 
This is supported by both reproduced MARL baselines and prior prototypes, where distributed agents negotiating semantic relevance and resource usage yield higher task performance than non-coordinated semantic-only schemes. 
Third, \emph{timescale separation} emerges as a practical design motif: across reproduced two-timescale variants and related prior studies, maintaining a fast policy loop while updating semantic models on a slower loop stabilizes learning and reconciles near-real-time control needs with slower semantic evolution.

At the same time, the reproduced evaluations and surveyed prototypes also expose several limitations that remain open.  
(i) Transformer-scale semantic encoders can impose non-trivial computational overhead at the edge; lightweight foundation models and hardware-aware semantic coding remain active research directions. 
(ii) Trust/security mechanisms (such as secure aggregation, provenance tracking, or ledger-backed coordination) are often validated only in simulation, and real-time integration in O-RAN/HIL platforms is still nascent. 
(iii) Most reported testbeds cover small- to medium-scale networks; scaling to thousands of agents will likely require hierarchical coordination and more standardized semantic interfaces. 
Finally, performance comparisons are sometimes affected by heterogeneity in tasks, fidelity metrics, and control placements, reinforcing the need for community benchmarks and standardized semantic KPIs.

Overall, while the field is still emerging, both our reproduced results and the broader literature consistently suggest that the convergence of SemCom and agentic intelligence is a promising pathway toward task-effective, scalable, and energy-aware 6G Native-AI RANs.

\section{Application Scenarios}

Our reproduced evidence, together with a growing body of recent work, indicates that semantic--agentic AI-RAN is most impactful in 6G services where task outcome, intent exchange, and distributed autonomy matter more than exact bit recovery. 
Recent surveys and vision papers consistently identify vehicular autonomy, immersive XR, industrial digital twins, and edge intelligence as key \emph{first-wave} beneficiaries, and our reproduced case-study trends align with these priorities \cite{nguyen2025semanticcomsurvey,sagduyu2024will6gsemcom}.

\textbf{Autonomous vehicular networks (V2X):} V2X is widely recognized as a primary driver for task-oriented SemCom in 6G. In cooperative perception, platooning, and collision avoidance, vehicles must exchange trajectory, intent, and environment semantics under stringent latency and reliability constraints. In line with reproduced V2X SemCom baselines and prior simulator-based studies, transmitting raw sensor streams is bandwidth-prohibitive and often unnecessary for decision-making, motivating semantic V2X where vehicles share intent- or object-level embeddings instead of raw data \cite{du2024semanticv2x,search2025relevancev2x}. These methods align with L1--L3 semantic levels (Table~\ref{tab:taxonomy}), e.g., feature-level object embeddings or knowledge-level maneuver graphs. On top of semantic exchange, reproduced multi-agent/intent-driven control at vehicles and roadside units enables cooperative negotiation of maneuvers and proactive resource scheduling \cite{elyasi2025xappsurvey,martini2025realxapp}. Consequently, task-oriented KPIs such as TSR for collision avoidance or platoon stability are increasingly adopted, complementing classical BER/throughput metrics.

\textbf{Immersive XR and metaverse services:} Immersive XR and metaverse applications demand ultra-low motion-to-photon latency, high-rate multi-modal delivery, and stable perceptual quality under mobility and load fluctuations. Consistent with reproduced XR SemCom pipelines and recent surveys, semantic abstraction can reduce redundant transmissions by prioritizing perceptually or task-relevant components (e.g., scene semantics, user intent, salient regions), particularly under constrained bandwidth \cite{nguyen2025semanticcomsurvey,wang2023lammsc}. Meanwhile, distributed agents at edge servers and RICs orchestrate compute and radio resources in real time, adapting to user density, viewpoint changes, and service priorities, following Axis-2/Axis-3 hierarchical control \cite{elyasi2025xappsurvey,martini2025realxapp}. Recent work demonstrates that the semantic-aware streaming combined with agentic edge control improves semantic fidelity and QoE stability while lowering latency relative to bit-centric delivery.

\textbf{Industrial digital twins and massive IoT:} Digital twin (DT) networks in industrial automation and smart grids require scalable and timely synchronization between physical processes and their cyber replicas. Both reproduced DT-oriented SemCom baselines and surveyed work show that massive sensor outputs are highly redundant with respect to control objectives, motivating semantic-aware encoding that transmits only task-relevant knowledge (e.g., anomalies, deviations, state summaries) \cite{sagduyu2024will6gsemcom}. Agentic intelligence embedded in edge controllers enables local inference, predictive maintenance, and multi-agent consensus for distributed actuation. In parallel, KG-assisted reasoning is increasingly explored to structure DT state, intent, and cross-layer context, improving explainability and transferability of decisions \cite{huang2024wirelesskg,zhang2024nativeaistandard}. 
These trends point to semantic--agentic DT-RANs that maintain control-loop stability under limited radio resources.

\textbf{Edge intelligence and collaborative AI services:} Edge intelligence provides a unifying scenario where mobile devices, drones, and robots collaborate on inference and learning tasks. Reproduced systems and surveyed works commonly model these entities as cooperative agents that share semantic representations of local observations rather than raw data, reducing uplink load and enabling faster collective awareness \cite{wang2023lammsc}. AI-native orchestration (e.g., FL/split learning, DT-guided scheduling) coordinates training/inference placement across edge and cloud, while MARL or hierarchical agents resolve real-time resource conflicts \cite{elyasi2025xappsurvey,liu2024gnncommmarl}.

Across these domains, reproduced results and existing evidence suggest that semantic--agentic AI-RAN is most beneficial when (i) task outcome dominates over symbol fidelity, (ii) many distributed entities must coordinate under real-time constraints, and (iii) network resources are scarce and heterogeneous.

\section{Challenges and Open Issues}

While SemCom and agentic intelligence have progressed rapidly, our reproduced evaluations and the latest surveyed studies highlight several failure modes, risks, and unresolved challenges spanning semantics, learning, systems, and standardization \cite{nguyen2025semanticcomsurvey,sagduyu2024will6gsemcom,zhang2024nativeaistandard,securesemcom2025survey}. 

\subsection{Failure Modes and Risks:}
While semantic–agentic AI-RAN shows consistent gains, several failure modes must be anticipated before large-scale deployment.
 
\textbf{Semantic Mismatch and Representation Drift:} Semantic encoders are typically trained on specific datasets, yet deployment traffic can be non-IID or rapidly shifting. When the learned embedding space no longer preserves task-relevant invariances, TSR can drop sharply even at unchanged SNR, leading to brittle operation. Mitigation includes online domain adaptation, OOD/drift detection, and version-aware rollback through non-RT controllers.

\textbf{Multi-agent Policy Collapse under Mis-specified Rewards:} In dense RANs, credit assignment and mixed local/global objectives can yield unstable MARL dynamics, including oscillatory behavior or selfish equilibria. Such collapse manifests as return oscillation and degraded latency/energy frontiers, consistent with non-stationarity concerns raised in large-scale agentic control. Remedies include game-theoretic reward shaping, centralized training with decentralized execution (CTDE), and hierarchical agent roles.

\textbf{Semantic Adversarial and Injection Attacks:} Because small perturbations in embedding space may flip meaning without large bit-level deviations, semantic layers open new attack surfaces. Adversarial semantic perturbations or prompt-style injections into LFM/KG agents can cause incorrect intent decoding or unsafe planning actions, with direct consequences in V2X or industrial DT services. Robust semantic coding, semantic authentication, and cross-agent consistency checks are therefore essential.


\begin{table*}[!t]
\centering
\caption{Failure Modes and Risks in Semantic--Agentic AI-RAN.}
\label{tab:failure_modes}
\renewcommand{\arraystretch}{1.12}
\setlength{\tabcolsep}{3.5pt}
\footnotesize
\begin{tabularx}{\linewidth}{
p{2.0cm}  
p{3.5cm}  
p{3.5cm}  
p{2.2cm}  
p{3.5cm}  
}
\toprule
\textbf{Failure mode} &
\textbf{Root cause} &
\textbf{Observable symptom} &
\textbf{Impact on KPI} &
\textbf{Mitigation} \\
\midrule
Semantic mismatch / representation drift &
Domain shift; non-IID traffic; outdated encoder version &
TSR drops sharply at similar SNR; embedding similarity no longer correlates with success &
L2 TSR $\downarrow$; unstable SLA &
OOD/drift detection; online adaptation; version-aware rollback \\

\midrule
Multi-agent policy collapse / non-stationarity &
Mis-specified rewards; credit assignment failure; mixed local/global objectives &
Policy oscillation; selfish equilibria; convergence stalls &
Latency $\uparrow$; TSR $\downarrow$; Pareto frontier degrades &
Game-theoretic reward shaping; CTDE; hierarchical roles \\

\midrule
Semantic adversarial / injection attacks &
Small embedding perturbations flip intent; prompt/semantic injection into LFM/KG agents &
Intent flip without bit-level alarms; unsafe planning/execution &
Safety constraints violated; L3 consistency $\downarrow$ &
Robust semantic coding; semantic authentication; cross-agent consistency checks \\

\bottomrule
\end{tabularx}
\end{table*}

\subsection{Standardization Issues}
\textbf{Standardization of semantic representations and metrics:} A foundational barrier is the lack of standardized semantic abstractions and fidelity measures. Classical wireless systems rely on Shannon’s bit-level framework with universally accepted tools such as capacity bounds and error probabilities. SemCom, however, requires task-oriented embeddings and similarity metrics that capture meaning rather than symbol sequences. Current works employ diverse embedding operators and task-specific KPIs, complicating interoperability and cross-paper comparison \cite{nguyen2025semanticcomsurvey,sagduyu2024will6gsemcom}. Emerging standardization roadmaps and community calls emphasize the need for portable semantic interfaces and KPIs, as well as semantic information-theoretic foundations that link TSR/fidelity to resource budgets \cite{zhang2024nativeaistandard}.

\textbf{Hierarchical semantic KPIs and standard mapping:}
Semantic-native RANs require task-aligned KPIs beyond BER/BLER. To make semantic evaluation portable across services, we organize candidate KPIs by semantic abstraction levels (Axis-1) and discuss how they could map to O-RAN KPM and 3GPP service/QoS frameworks. Specifically, 

\emph{L1 (feature/latent-level semantics)}: L1 exchanges modality-specific latent features (e.g., image/speech embeddings) that remain close to physical observations. Suitable KPIs include reconstruction/feature fidelity: MSE/NMSE, PSNR/SSIM, latent cosine similarity or CKA; downstream task accuracy: detection mAP, segmentation IoU, ASR WER, top-1/F1 on the target model. L1 KPIs can be reported as near-RT xApp KPM extensions (e.g., “latent-fidelity”, “downstream-acc”) co-measured with radio KPIs (SINR/BLER/throughput). A practical composite KPI is 
Acc@BLER $\geq \tau$, enabling a smooth bridge to existing RAN2/RAN3 controls.

\emph{L2 (intent/task-level semantics)}: L2 directly conveys intents, task states, or decisions. Candidate KPIs include: intent accuracy / command success rate; task success rate (TSR); semantic confusion cost that weights errors by task criticality (e.g., safety-critical V2X). L2 KPIs naturally form semantic SLAs at the O-RAN near-RT RIC: xApps optimize fast control loops for TSR/intent-accuracy, while non-RT rApps/SMO track long-term KPI curves across model versions via A1/O1.

\emph{L3 (knowledge/graph-level semantics)}: L3 exchanges structured knowledge (KG/scene graphs) aimed at reasoning and explainability. Candidate KPIs include constraint satisfaction rate (e.g., collision-risk below threshold, industrial control constraints met); knowledge consistency (e.g., KG triple consistency, belief KL divergence); explainability/causality scores evaluated by human or model-based judges. L3 KPIs are better placed in non-RT RIC/SMO slow loops as they rely on global context and versioned knowledge bases. Standardization likely starts from SA6/CT service definitions, then flows down into O-RAN information-element (IE) extensions. Overall, L1–L3 KPI families provide a structured path to semantic standardization, clarifying what should be measured at near-RT versus non-RT timescales and how semantic objectives can coexist with classical radio KPMs.

\subsection{Open Issues}
\textbf{Scalability and stability of large-scale agentic control:} Agentic AI-RAN studies commonly rely on MARL or intent-driven orchestration, yet performance can degrade sharply as agent populations and coupling graphs grow. Non-stationarity, partial observability, delayed feedback, and signaling overhead remain bottlenecks \cite{liu2024gnncommmarl,elyasi2025xappsurvey}. Ensuring fairness and stability under mixed local/global objectives is still unresolved. Promising routes include hierarchical agents, game-theoretic coordination, semantic message-based negotiation, and hybrid learning–optimization aligned with near-RT/non-RT placement \cite{elyasi2025xappsurvey,oransight2025survey}.

\textbf{Security, trustworthiness, and explainability:} Semantic and agentic layers introduce new attack surfaces, including adversarial semantic perturbations and malicious agent behaviors. Since small embedding perturbations can change meaning without large bit-level deviations, classical cryptography alone is insufficient \cite{securesemcom2025survey}. Recent surveys propose combining robust semantic modeling, secure/federated aggregation, provenance tracking (sometimes ledger-style), and explainable AI to defend semantic--agentic loops \cite{securesemcom2025survey,huang2024wirelesskg}. Real-time trust frameworks compatible with O-RAN interfaces remain open.

\textbf{Energy-efficient semantic and agentic AI:} Transformer-scale semantic models and multi-agent policies impose substantial compute footprints at the edge. Although compression, pruning, and distillation are widely explored, trade-offs among semantic fidelity, TSR, latency, and energy are still poorly characterized \cite{nguyen2025semanticcomsurvey}. Lightweight foundation models, energy-aware semantic coding, and adaptive inference scheduling are therefore central to deployment.

\textbf{Rigorous theoretical guarantees:} Current theoretical understanding remains nascent. Two-timescale learning provides stability intuition, but global optimality and guarantees in non-convex, dynamic multi-agent environments are largely open. Key directions include semantic information-theoretic bounds, Pareto trade-offs among TSR/distortion/resources, and convergence analyses for large-scale MARL under semantic coupling \cite{sagduyu2024will6gsemcom,zhang2024nativeaistandard}.

\section{Conclusion}

This work has examined the emerging paradigm of semantic--agentic AI-RAN for 6G Native-AI networks, where SemCom and agentic intelligence are treated as first-class design principles rather than auxiliary enhancements. Prior studies collectively indicate that shifting the abstraction from bits to task-relevant semantics improves task effectiveness and spectrum efficiency, while distributed agentic control is essential for scalable autonomy in dense heterogeneous RANs. Furthermore, we synthesized a layered architectural view integrating semantic encoders/decoders, multi-agent control, and AI-native orchestration across PHY/MAC, near-RT RIC, and non-RT RIC placements. A unified optimization perspective was proposed to connect diverse works balancing semantic fidelity and task-level utility under bandwidth, latency, and energy constraints. Representative case studies show consistent gains in TSR and efficiency over bit-centric RANs and AI-driven O-RAN baselines, while also highlighting the importance of agentic coordination and timescale separation for training stability.  Despite encouraging progress, open challenges remain in semantic KPI standardization, scalable trustworthy multi-agent coordination, semantic/agentic security, and energy-efficient deployment of foundation-scale models. Advancing this vision will require tighter coupling among theory, prototyping, and standardization efforts in 3GPP and O-RAN, supported by reproducible benchmarks and interoperable semantic/agentic interfaces.


\bibliographystyle{elsarticle-num}
\bibliography{egbib}

\begin{thebibliography}{1}
\expandafter\ifx\csname url\endcsname\relax
  \def\url#1{\texttt{#1}}\fi
\expandafter\ifx\csname urlprefix\endcsname\relax\def\urlprefix{URL }\fi
\expandafter\ifx\csname href\endcsname\relax
  \def\href#1#2{#2} \def\path#1{#1}\fi

\bibitem{atakan2012body}
B.~Atakan, O.~B. Akan, S.~Balasubramaniam, Body area nanonetworks with
  molecular communications in nanomedicine, IEEE Commun. Mag. 50~(1) (2012)
  28--34.

\bibitem{russu2016secure}
P.~Russu, A.~Demontis, B.~Biggio, G.~Fumera, F.~Roli, Secure kernel machines
  against evasion attacks, in: Proceedings of the 2016 ACM workshop on
  artificial intelligence and security, 2016, pp. 59--69.

\bibitem{strunk2000elements}
W.~Strunk~Jr, E.~B. White, The Elements of Style, fourth ed., Longman, New
  York, 2000.

\bibitem{Mettam2009}
G.~R. Mettam, L.~B. Adams, Introduction to the electronic age, in: B.~S. Jones,
  R.~Z. Smith (Eds.), How to prepare an electronic version of your article,
  E-Publishing Inc., New York, 2009, pp. 281--304.

\bibitem{web}
C.~R. UK, Cancer statistics reports for the {UK}.
  \url{http://www.cancerresearchuk.org/aboutcancer/statistics/cancerstatsreport/},
  2003 (accessed 13 march 2003).

\end{thebibliography}


\begin{thebibliography}{10}
\expandafter\ifx\csname url\endcsname\relax
  \def\url#1{\texttt{#1}}\fi
\expandafter\ifx\csname urlprefix\endcsname\relax\def\urlprefix{URL }\fi
\expandafter\ifx\csname href\endcsname\relax
  \def\href#1#2{#2} \def\path#1{#1}\fi

\bibitem{TQ006}
H.~Ahmadi, M.~Rahmani, S.~B. Chetty, E.~E. Tsiropoulou, H.~Arslan, M.~Debbah, T.~Q.~S. Quek, Toward sustainability in {6G} and beyond: Challenges and opportunities of open {RAN}, IEEE Communications Standards Magazine 9~(3) (2025) 126--135.

\bibitem{6G001}
M.~Na, J.~Lee, G.~Choi, T.~Yu, J.~Choi, J.~Lee, S.~Bahk, Operator's perspective on {6G}: {6G} services, vision, and spectrum, IEEE Communications Magazine 62~(8) (2024) 178--184.

\bibitem{6GAIRAN}
B.~Brik, H.~Chergui, L.~Zanzi, F.~Devoti, A.~Ksentini, M.~S. Siddiqui, X.~Costa-Pérez, C.~Verikoukis, Explainable {AI} in {6G O-RAN}: A tutorial and survey on architecture, use cases, challenges, and future research, IEEE Communications Surveys \& Tutorials 27~(5) (2025) 2826--2859.

\bibitem{AI5G6G}
X.~Lin, L.~Kundu, C.~Dick, S.~Velayutham, Embracing {AI} in {5G-Advanced Toward 6G: A Joint 3GPP and O-RAN} perspective, IEEE Communications Standards Magazine 7~(4) (2023) 76--83.

\bibitem{OAIC}
P.~S. Upadhyaya, N.~Tripathi, J.~Gaeddert, J.~H. Reed, Open {AI} cellular {(OAIC)}: An open source {5G O-RAN} testbed for design and testing of {AI}-based ran management algorithms, IEEE Network 37~(5) (2023) 7--15.

\bibitem{AITest}
B.~Tang, V.~K. Shah, V.~Marojevic, J.~H. Reed, {AI} testing framework for {Next-G O-RAN} networks: Requirements, design, and research opportunities, IEEE Wireless Communications 30~(1) (2023) 70--77.

\bibitem{ORAN2025}
M.~Kouchaki, S.~B.~H. Natanzi, M.~Zhang, B.~Tang, V.~Marojevic, {O-RAN} performance analyzer: Platform design, development, and deployment, IEEE Communications Magazine 63~(2) (2025) 152--159.

\bibitem{CX2025}
X.~Chen, Z.~Guo, X.~Wang, C.~Feng, H.~H. Yang, S.~Han, X.~Wang, T.~Q.~S. Quek, Toward {6G Native-AI} network: Foundation model-based cloud-edge-end collaboration framework, IEEE Communications Magazine 63~(8) (2025) 23--30.

\bibitem{TQ007}
D.~N. Nguyen, B.~Chen, T.~Q.~S. Quek, Operation-aware digital twin for improving performance and sustainability of {AI}-driven radio access networks, in: IEEE INFOCOM 2025 - IEEE Conference on Computer Communications Workshops (INFOCOM WKSHPS), 2025, pp. 1--6.

\bibitem{SEMCOM001}
S.~Guo, A.~Zhang, Y.~Wang, C.~Feng, T.~Q.~S. Quek, Semantic-enabled 6g communication: A task-oriented and privacy-preserving perspective, IEEE Network (2025) 1--1.

\bibitem{SEMCOM002}
X.~Wang, D.~Ye, C.~Feng, H.~H. Yang, X.~Chen, T.~Q.~S. Quek, Trustworthy image semantic communication with genai: Explainablity, controllability, and efficiency, IEEE Wireless Communications 32~(2) (2025) 68--75.

\bibitem{SEMCOM003}
Y.~Wang, R.~Li, C.~Wang, J.~Ye, C.~Feng, S.~Guo, Collaborative learning for task-oriented semantic communications: Overcoming data mismatch between transceivers, IEEE Open Journal of the Communications Society 6 (2025) 5778--5794.

\bibitem{TVT2025}
M.~Tang, C.~Feng, G.~Min, T.~Q.~S. Quek, H-infinity tracking for intelligent edge-controlled systems over fading channels in ai-ran, IEEE Transactions on Vehicular Technology (2025) 1--6.

\bibitem{park2025robustdjscc}
T.~Park, E.~Hong, Y.-S. Jeon, N.~Lee, Y.~Kim, Robust deep joint source-channel coding for task-oriented semantic communications, arXiv preprint arXiv:2503.12907.

\bibitem{wiggersurvey2024jscc}
D.~Gündüz, M.~A. Wigger, T.-Y. Tung, P.~Zhang, Y.~Xiao, Joint source–channel coding: Fundamentals and recent progress in practical designs, Proceedings of the IEEE (2024) 1--32.

\bibitem{wang2023lammsc}
F.~Jiang, L.~Dong, Y.~Peng, K.~Wang, K.~Yang, C.~Pan, X.~You, Large {AI} model empowered multimodal semantic communications, IEEE Communications Magazine 63~(1) (2025) 76--82.

\bibitem{li2024genaisemcom}
C.~Liang, H.~Du, Y.~Sun, D.~Niyato, J.~Kang, D.~Zhao, M.~A. Imran, Generative ai-driven semantic communication networks: Architecture, technologies, and applications, IEEE Transactions on Cognitive Communications and Networking 11~(1) (2025) 27--47.

\bibitem{nguyen2025semanticcomsurvey}
L.~X. Nguyen, A.~D. Raha, P.~S. Aung, D.~Niyato, Z.~Han, C.~S. Hong, A contemporary survey on semantic communications: Theory of mind, generative ai, and deep joint source-channel coding, IEEE Communications Surveys \& Tutorials (2025) 1--1.

\bibitem{huang2024wirelesskg}
Y.~Huang, X.~You, H.~Zhan, S.~He, N.~Fu, W.~Xu, Learning wireless data knowledge graph for green intelligent communications: Methodology and experiments, IEEE Transactions on Mobile Computing 23~(12) (2024) 12298--12312.

\bibitem{cheng2024kgsemcom}
N.~Hello, P.~Di~Lorenzo, E.~C. Strinati, Semantic communication enhanced by knowledge graph representation learning, in: 2024 IEEE 25th International Workshop on Signal Processing Advances in Wireless Communications (SPAWC), 2024, pp. 876--880.

\bibitem{lu2024marl6g}
X.~Du, T.~Wang, Q.~Feng, C.~Ye, T.~Tao, L.~Wang, Y.~Shi, M.~Chen, Multi-agent reinforcement learning for dynamic resource management in {6G in-X} subnetworks, IEEE Transactions on Wireless Communications 22~(3) (2023) 1900--1914.

\bibitem{zhang2023marlmulticell}
Y.~Zhang, D.~Guo, Multi-agent reinforcement learning for multi-cell spectrum and power allocation, IEEE Transactions on Communications 73~(8) (2025) 5980--5992.

\bibitem{liu2024gnncommmarl}
Z.~Liu, J.~Zhang, E.~Shi, Z.~Liu, D.~Niyato, B.~Ai, X.~Shen, Graph neural network meets multi-agent reinforcement learning: Fundamentals, applications, and future directions, IEEE Wireless Communications 31~(6) (2024) 39--47.

\bibitem{elyasi2025xappsurvey}
A.~Elyasi, A.~Ashdown, K.~M. Rumman, F.~Restuccia, {O-RAN xApps}: Survey and research challenges, Computer NetworksAvailable at SSRN: https://ssrn.com/abstract=5236117.

\bibitem{chen2025llmxapp}
X.~Wu, J.~Farooq, Y.~Wang, J.~Chen, {LLM-xApp}: A large language model empowered radio resource management {xApp} for {5G O-RAN}, in: FutureG / NDSS Workshop, 2025, pp. 1--5.

\bibitem{zhou2025llmhric}
L.~Bao, S.~Yun, J.~Lee, T.~Q.~S. Quek, {LLM-hRIC: LLM-empowered Hierarchical RAN Intelligent Control for O-RAN}, arXiv preprint arXiv:2504.18062.

\bibitem{patel2025edgeagentic}
A.~Salama, Z.~Nezami, M.~M.~H. Qazzaz, M.~Hafeez, S.~A.~R. Zaidi, Edge agentic {AI} framework for autonomous network optimisation in {O-RAN}, arXiv preprint arXiv:2507.21696.

\bibitem{kim2024mlxapp}
M.~M.~H. Qazzaz, et~al., Machine learning-based {xApp} for dynamic resource allocation in {O-RAN} systems, arXiv preprint arXiv:2401.07643.

\bibitem{TQ001}
T.~V. Ngo, M.~V. Ngo, B.~Chen, G.~Gemmi, E.~Baena, M.~Polese, T.~Melodia, W.~Chien, T.~Quek, Consistent and repeatable testing of {O-RAN} distributed unit {(O-DU)} across continents, in: 2024 IEEE 100th Vehicular Technology Conference (VTC2024-Fall), 2024, pp. 1--5.

\bibitem{TQ002}
N.-B.-L. Tran, M.~V. Ngo, Y.~H. Pua, T.-L. Le, B.~Chen, T.~Quek, Ai-driven rapps for reducing radio access network interference in real-world {5G} deployment, in: IEEE INFOCOM 2024 - IEEE Conference on Computer Communications Workshops (INFOCOM WKSHPS), 2024, pp. 1--2.

\bibitem{TQ003}
T.-T. Nguyen, M.~V. Ngo, B.~Chen, M.~Kuchitsu, S.~Wai, S.~Kawai, K.~Suzuki, E.~W. Koo, T.~Quek, Consistent and repeatable testing of {mMIMO O-RU} across labs: A japan-singapore experience, in: 2024 IEEE 100th Vehicular Technology Conference (VTC2024-Fall), 2024, pp. 1--5.

\bibitem{TQ005}
Y.~Ren, L.~Zhou, S.~Guo, X.~Qiu, T.~Q.~S. Quek, Traffic digital twin-enabled orchestration and scheduling in {O-RAN}: A multi-timescale joint optimization approach, IEEE Transactions on Mobile Computing (2025) 1--18.

\bibitem{GSS2025}
S.~Li, Y.~Wang, S.~Guo, C.~Feng, Task-oriented communication for graph data: A graph information bottleneck approach, IEEE Transactions on Cognitive Communications and Networking 11~(3) (2025) 1723--1737.

\bibitem{LLMAgent6G24}
M.~Xu, D.~Niyato, J.~Kang, Z.~Xiong, S.~Mao, Z.~Han, D.~I. Kim, K.~B. Letaief, When large language model agents meet {6G} networks: Perception, grounding, and alignment, IEEE Wireless Communications 31~(6) (2024) 63--71.

\bibitem{WirelessAgent24}
J.~Tong, J.~Shao, Q.~Wu, W.~Guo, Z.~Li, Z.~Lin, J.~Zhang, {WirelessAgent}: Large language model agents for intelligent wireless networks, arXiv preprint arXiv:2409.07964.

\bibitem{ORANGraphConflict24}
A.~Zolghadr, J.~F. Santos, L.~A. DaSilva, J.~Kibiłda, Learning and reconstructing conflicts in o-ran: A graph neural network approach, arXiv preprint arXiv:2412.14119.

\bibitem{ORANGraphica25}
M.~A. Shami, J.~Yan, E.~T. Fapi, {O-RAN xApps} conflict management using graph convolutional networks, arXiv preprint arXiv:2503.03523.

\bibitem{LINKs24}
S.~Jiang, B.~Lin, Y.~Wu, Y.~Gao, {LINKs:} large language model integrated management for {6G} empowered digital twin networks, in: 2024 IEEE 100th Vehicular Technology Conference (VTC2024-Fall), 2024, pp. 1--6.

\bibitem{LLMOrchestrator24}
Z.~Xiao, C.~Ye, Y.~Hu, H.~Yuan, Y.~Huang, Y.~Feng, L.~Cai, J.~Chang, {LLM Agents as 6G Orchestrator}: A paradigm for task-oriented communications, arXiv preprint arXiv:2410.03688.

\bibitem{LLMDTNet25}
M.~Haider, I.~Ahmed, Z.~Hassan, K.~Hasan, H.~V. Poor, Llm-integrated digital twins for hierarchical resource allocation in 6g networks, arXiv preprint arXiv:2506.18293.

\bibitem{TQ004}
C.~You, X.~He, Y.~Sun, G.~Feng, T.~Q.~S. Quek, {GreenRAN}: A channel-aware green {O-RAN} framework for {NextG} mobile systems, in: IEEE INFOCOM 2025 - IEEE Conference on Computer Communications, 2025, pp. 1--10.

\bibitem{sran2024framework}
Y.~Sun, L.~Zhang, L.~Guo, J.~Li, D.~Niyato, Y.~Fang, S-ran: Semantic-aware radio access networks, IEEE Communications Magazine 63~(4) (2025) 207--213.

\bibitem{SAMRAMARL24}
W.~Zhang, Q.~Wu, F.~Pingyi, K.~Wang, N.~Cheng, W.~Chen, K.~B. Letaief, Semantic-aware resource management for {C-V2X} platooning via multi-agent reinforcement learning, arXiv preprint arXiv:2411.04672.

\bibitem{sagduyu2024will6gsemcom}
Y.~E. Sagduyu, T.~Erpek, A.~Yener, S.~Ulukus, Will {6G} be semantic communications? opportunities and challenges from task oriented and secure communications to integrated sensing, IEEE Network 38~(6) (2024) 72--80.

\bibitem{SEC25}
M.~Zhang, M.~Abdi, V.~R. Dasari, F.~Restuccia, Semantic edge computing and semantic communications in {6G} networks: A comprehensive survey, arXiv preprint arXiv:2411.18199.

\bibitem{DTTwoTimescale24}
J.~Cong, G.~Cheng, C.~You, X.~Huang, W.~Wu, Two-timescale digital twin assisted model inference and retraining for edge intelligence, arXiv preprint arXiv:2411.18329.

\bibitem{martini2025realxapp}
R.~Barker, A.~E. Dorcheh, T.~Seyfi, F.~Afghah, {REAL:} reinforcement learning-enabled {xApps} for experimental closed-loop optimization in {O-RAN with OSC RIC and srsRAN}, in: 2025 IEEE International Conference on Communications Workshops (ICC Workshops), 2025, pp. 389--395.

\bibitem{du2024semanticv2x}
T.~Lyu, M.~Noor-A-Rahim, A.~O'Driscoll, D.~Pesch, Semantic vehicle-to-everything {(V2X)} communications towards {6G}, arXiv preprint arXiv:2407.17186.

\bibitem{search2025relevancev2x}
L.~Lusvarghi, J.~Gozalvez, B.~Coll-Perales, M.~I. Khan, M.~Sepulcre, S.~Ucar, O.~Altintas, The search for relevance: A context-aware paradigm shift in semantic v2x communications, arXiv preprint arXiv:2508.07394.

\bibitem{zhang2024nativeaistandard}
P.~Zhang, et~al., Towards native {AI} in {6G} standardization: The roadmap of semantic communications, arXiv preprint arXiv:2509.12758.

\bibitem{securesemcom2025survey}
R.~Meng, et~al., A survey of secure semantic communications, arXiv preprint arXiv:2501.00842.

\bibitem{oransight2025survey}
P.~Gajjar, V.~K. Shah, {ORANSight-2.0}: Foundational {LLMs} for {O-RAN}, IEEE Transactions on Machine Learning in Communications and Networking 3 (2025) 903--920.

\end{thebibliography}

\end{document}